\documentclass[fleqn,10pt]{wlscirep}
\usepackage[utf8]{inputenc}
\usepackage[T1]{fontenc}
\usepackage{braket}
\usepackage{comment}
\usepackage{float}
\title{Active Learning Guided Computational Discovery of 2D Materials with Large Spin Hall Conductivity 
}

\author[1,2,$\dagger$]{Abhijeet J. Kale}
\author[1,$\dagger$]{Sanjeev S. Navaratna}
\author[2,3,$\dagger$]{Pratik Sahu}
\author[4]{Henry Chan}
\author[2,3,*]{B. R. K. Nanda}
\author[1,2,*]{Rohit Batra}

\affil[1 ]{ Materials Informatics Lab, Department of Metallurgical and Materials Engineering, Indian Institute of Technology Madras, Chennai 600036, India}
\affil[2 ]{ Center for Atomistic Modelling and Materials Design, Indian Institute of Technology Madras, Chennai 600036, India}
\affil[3 ]{ Condensed Matter Theory and Computational Lab, Department of Physics, Indian Institute of Technology Madras, Chennai 600036, India}
\affil[4 ]{ Center for Nanoscale Materials, Argonne National Laboratory, Lemont, IL, USA}
\affil[$\dagger$ ]{ equal contribution}
\affil[* ]{ Corresponding authors: nandab@iitm.ac.in and rbatra@smail.iitm.ac.in.}

\begin{abstract} 
Two-dimensional (2D) materials are promising candidates for next-generation spintronic devices due to their tunable properties and potential for efficient spin-charge interconversion. However, discovering materials with intrinsically high spin Hall conductivity (SHC) is hindered by the vast chemical space and expensive nature of conventional experimental and first-principles methods. In this work, we employ an active learning framework to accelerate the discovery of high-SHC 2D materials. Machine learning (ML)  models were trained on SHC values computed from density functional theory calculations, incorporating the Kubo formalism via tight-binding Hamiltonians constructed from maximally localized Wannier functions, with explicit treatment of spin-orbit coupling. Starting from random but chemically diverse 24 2D systems, the dataset was expanded to 41 cases (from an overall pool of \textasciitilde 2000 materials) over three active learning loops using an expected improvement acquisition strategy. The ML technique successfully identified several high SHC candidates with the best candidate exhibiting a SHC of 271.52 $\hbar$/e $\Omega$\textsuperscript{-1}, nearly 23 times higher than the top performer in the initial round. Beyond candidate discovery, several features such as orbital symmetry near the Fermi energy, types of atomic species, material composition, covalent radii, and electronegativity of constituent atoms were found to play critical role in shaping the spin Hall response in 2D systems. The data generated is made publicly available to facilitate further advances in 2D spintronics.
\end{abstract}
\begin{document}

\flushbottom
\maketitle

\thispagestyle{empty}

\section*{INTRODUCTION}
Spintronics devices manipulate spin degree of freedom of electrons to process, transport, compute, store information, and generate signals. %Underlying mechanism
The deterministic factor for many spintronic technologies is the spin Hall effect (SHE), wherein longitudinal charge current (I\textsubscript{c}) is converted into transverse spin current (I\textsubscript{s}) via intrinsic spin-orbit coupling (SOC)\cite{sinova_spin_2015}. This paves the way for easier utilization of non-magnetic materials in the spintronic industry. Importantly, the generated pure spin flow, I\textsubscript{s}, does not involve flow of net charge and thus energy efficient information processing can be realized without Joule heating losses \cite{li_first-principles_2016}. Moreover, upon diffusion of I\textsubscript{s} into an adjacent ferromagnetic (FM) layer, its angular momentum exerts a torque to either oscillate or switch magnetization of the FM layer, enabling spin-orbit torque functionalities such as non-volatile data storage and processing, spin logic devices, and neuromorphic computing, among others.
%THz source and spectroscopy, etc.
\cite{ren_hybrid_2023}
%Spin flow can also occur with spin polarized current but it accompanies bodily motion of charges that delimits above advantages. The pure spin current can be generated via SHE and spin pumping.\cite{takahashi_physical_2016}
%The efficiency of I\textsubscript{c} to I\textsubscript{s} conversion is quantified by a metric called spin Hall conductivity (SHC) tensor defined as the density of generated I\textsubscript{s} per unit of applied electric field. 

Extensive computational and experimental efforts have been dedicated to discovering materials with high SHC. Chronologically, this search has so far focused on 3D bulk materials such as heavy 5\textit{d} transition metals \cite{kimura_room-temperature_2007, guo_ab_2009,pai_spin_2012}, GaAs \cite{chernyshov_evidence_2009}, and bulk transition metal dichalcogenides (TMDs) \cite{xu_high_2020, sahu_emergence_2024}. %However, the (higher) crystal symmetries in 3D bulk cases can impose restrictions on the forms of physical properties of the crystal (such as SHC) according to Neumann's principle.\cite{xiang_classification_2023} 
Over the last decade, the emerging class of 2D materials has demonstrated promising spin-driven quantum phenomena and hence form a natural platform to investigate SHE \cite{ahn_2d_2020, wunderlich_experimental_2005, song_coexistence_2020}. Furthermore, for efficient and scalable spintronic devices, materials are expected to show significant spin-related properties such as spin diffusion length, spin relaxation time, and spin coherence. %Efforts have been made to break the crystal symmetry of 3D materials by down-scaling them into 2D structures as it can lift the trade-off between spin-related properties, enabling 2D system to function as both spin generators and transporters, and facilitating different forms of SHE.\cite{song_coexistence_2020}  
The flagship 2D material - graphene - is theoretically estimated to exhibit spin diffusion length and relaxation time a few orders of magnitude higher than traditional conductors \cite{han_graphene_2014}. However, its direct usage as spin channel or Hall layer is limited due to weak SOC and as a poor host of FM state. %and inability of 2D lattice to host FM state as predicted by the Mermin-Wagner theorem. \cite{mermin_absence_1966} 
Thus, efforts have been made to introduce FM state in graphene via doping\cite{hong_room-temperature_2012,nair_dual_2013}, introducing vacancies\cite{yazyev_defect-induced_2007, cervenka_room-temperature_2009, nanda_electronic_2012, padmanabhan_intertwined_2016}, and through magnetic proximity\cite{haugen_spin_2008, yang_proximity_2013}. Another widely adopted research direction in 2D spintronics is the coupling of 2D TMDs (with high SOC and broken inversion symmetry) with FM alloy \cite{shao_strong_2016, mudgal_magnetic-proximity-induced_2023} or FM van der Waals materials \cite{dai_interfacial_2024} in the form of heterostructures. Huge chemical diversity, large SOC due to more confined and correlated electrons \cite{meitei_electron_2024, jadaun_rational_2020, soumyanarayanan_emergent_2016}, bandgap tunability via external factors (electric field, strain, etc.), and easier integration with present industrial infrastructure \cite{kong_path_2019} are some reasons why 2D materials are generally considered promising for spintronic applications.
%2D materials offer adaptability to numerous heterostructure configurations, easier integration with industrial setups\cite{kong_path_2019}, robust SOC-based effect at low dimensions \cite{soumyanarayanan_emergent_2016}, SOC and bandgap tunability via external routes (electric field, strain, proximity effect), mechanical flexibility, etc. to realize space-efficient spintronics technologies. 

\par The SHC is influenced by both intrinsic and extrinsic mechanisms. The intrinsic contribution arises from the SOC-influenced electronic band structure (EBS), while the extrinsic contribution originates from scattering processes due to impurities, defects, and other forms of disorder. Notably, the intrinsic SHC can be estimated reliably using DFT with incorporation of SOC\cite{guo_intrinsic_2008, sagasta_tuning_2016}. 
%SHC contributed by scattering due to disorder (extrinsic) and electronic bands influenced by SOC (intrinsic). It is relatively less complex and reasonably accurate to predict intrinsic SHC with computational techniques.\cite{guo_intrinsic_2008, sagasta_tuning_2016}
The SOC splits and shifts the band to give non-zero Berry curvature, %$\Omega_{n}^{z}$%
 which serves as a proxy magnetic field in the momentum space. Such split can differentiate the spin population leading to a net spin current \cite{sahu_emergence_2024, mudgal_magnetic-proximity-induced_2023}. Motivated by the search for materials with large SHC, several high-throughput computational studies have evaluated the intrinsic SHC across a broad range of bulk materials \cite{zhang_different_2021, ji_spin_2022, xu_high-throughput_2024}. More recently, similar efforts have been extended to 2D systems, including a dataset of 426 rare-earth-free monolayers \cite{zhou_high-throughput_2025}.
%but was limited to 6 atoms per unit cell.\cite{zhou_high-throughput_2025} However, 2D systems exist beyond rare-earth elements, can have more number of atoms per unit cell and are more than couple of thousands in numbers. 
\textcolor{black}{While such high-throughput approaches are invaluable for large-scale screening, their accuracy may be compromised due to algorithmic simplifications and coarse approximations made to accelerate the calculations.} For instance, a crucial aspect of SHC computations is the construction of tight-binding (TB) models using maximally localized Wannier functions (MLWFs) derived from \textit{ab initio} calculations \cite{qiao_calculation_2018}. In high-throughput workflows, the generation of MLWFs is typically automated, \cite{vitale_automated_2020, garrity_database_2021, gresch_automated_2018} %\textcolor{red}{or done using random projections},
a process which may be prone to errors particularly for dispersive band structures, where it is important to correctly identify orbitals to be considered for projections. Thus, the quality of wannierization can significantly affect the accuracy of SHC computations. %\textcolor{red}{Similarly, the efficiency of high-throughput workflows is also limited since the decision of which material is to be studied is random (or sometimes exhaustive), leading to substantial computational effort being spent on modeling materials systems with little potential for high SHC (not necessarily correct and should be deleted)}.
%Moreover, in computational SHC studies, the local description of EBS based on Wannier functions is utilized to construct tight-binding (TB) models.\cite{qiao_calculation_2018} In several studies on screening candidates from a larger pool, the onerous task of generating maximally localized Wannier functions (MLWFs) or corresponding TB Hamiltonians is delegated to the automated high-throughput methods.\cite{vitale_automated_2020, garrity_database_2021, gresch_automated_2018} However, the accuracy of automation can be limited for the properties which strongly depends on the specific symmetry of MLWFs or can be hampered by compromise on Wannierization quality (dispersive bandstructures).

To overcome the limitations of the existing computational approaches and to efficiently screen 2D materials with high SHC values, in this work, we develop an active learning framework that combines DFT calculations, TB modeling and ML methods. 
%Thus, existing computational approaches are limited in either their efficiency (costlier in terms of resources, time and quantitative output) or in their accuracy (due to automation). Given the vastness of 2D materials, there is a need to discover materials with high SHC and establish 2D SHC database to test the spintronic potential of various configurations. Therefore, to overcome the limitations mentioned above, we report the use of active ML framework to guide the search for high SHC materials from a large database of nearly two thousands 2D materials.
The ML models uses the concept of expected improvement to recommend candidates from a large pool of nearly 2000 2D materials with potential for high SHC. Selected candidates are then further evaluated using combined DFT and TB modeling approach, incorporating user-guided construction of MLWF to ensure accurate SHC evaluation. The complete process is carried iteratively in an active learning framework with around 5-7 candidates examined in each round using the aforementioned electronic structure methods. Overall, across three active learning cycles, we investigated 41 materials and observed a consistent improvement in SHC values in each round, culminating in the discovery of several novel compounds with SHC significantly exceeding the known materials. Notably, the top-performing material, Fe\textsubscript{2}TeSe, was found to have SHC nearly 23 times greater than the best candidate in the initial round.
%we studied a total of 41 2D materials across three active learning loops and successfully found materials with higher and higher SHC values in each round. Importantly, we found many novel candidates with SHC much larger than the known 2D materials, with the best candidate (Fe2TeSe) having SHC value almost 23 times higher than the best initial candidate.
The SHC data and the associated ML models were also analyzed to extract key chemical trends that result in high SHC. Contribution of \textit{d}- electrons near to the Fermi energy (E\textsubscript{F}), difference in the covalent radius, and the number of mirror symmetry planes are some example factors that were found to significantly impact the SHC. Overall, this work not only led to successful identification of 2D materials with high computed SHC, but it also shows how materials informatics approaches can be combined with accurate but expensive physics-based models to efficiently search for materials from large chemical spaces.

\section*{RESULTS and DISCUSSION} Figure 1 shows the schema of our workflow that incorporates  DFT, empirical TB modeling and data-driven ML method to discover 2D materials with high SHC. Overall, the role of ML methods is to learn the complex relationship between SHC and fundamental features of materials emerging from their crystal symmetry and electronic structure of individual elements, and recommend potential candidates expected to exhibit large SHC from a large 2D materials database. The overall search for 2D spintronic materials with high SHC is partitioned into three parts. First, the EBS of a 2D material taken from cleaned version of Materials Cloud Two-Dimensional Structure Database (MC2D)  \cite{mounet_two-dimensional_2018, campi_expansion_2023} is calculated using DFT. As the second step, the minimal basis set based TB modeling is then adopted using MLWF formalism to compute the Berry curvature and thereby the SHC using the Kubo formalism. %once the EBS of crystalline systems is described in terms of localized orbitals instead of \textit{k}-space extended Bloch states \(\ket{\psi_{nk}}\). For this, projection matrices $A_{mn}^{k}$ generated from user-specified target Bloch bands (\textit{m}) and atomic orbitals (\textit{n}) as well as overlap matrices $M_{mn}^{k,b}$ derived from periodic part of \(\ket{\psi_{nk}}\) are used to construct the real-space MLWFs \(\ket{w_{nR}}\). %Here, \textit{R} is a real space lattice vector and \textit{b} is a displacement vector in reciprocal \textit{k} space. MLWFs provide an alternate EBS representation wherein a band comprised of complex plane-waves is represented by a single function.
%The TB model Hamiltonian \(\mathcal{H}_{\text{TB}}\), with SOC corrections, is next constructed to compute the Berry curvature $\Omega_{n}^{z}$ and derive SHC using the Kubo formalism. 
%The spin-orbit coupling (SOC) is separately treated in EBS and later introduced to TB model Hamiltonian \(\mathcal{H}_{\text{TB}}\). Based on MLWFs with corresponding hopping parameters \textit{t\textsubscript{nn'}} and SOC corrections, the \(\mathcal{H}_{\text{TB}}\) calculates Berry curvature $\Omega_{n}^{z}$ to compute SHC using Kubo formalism thereby completing the second part.
The third part comprises of leveraging ML techniques to train models on the dataset of computed SHC values, and screening potential candidates expected to exhibit large SHC from a vast pool of 2D systems. To achieve this, we first created an initial training dataset of 24 diverse 2D systems and computed their SHC values. A range of hand-crafted symmetry features as well as elemental features from standard \textit{matminer} \cite{ward_matminer_2018} library were then used to train a series of traditional  regression models on this dataset. Since the range of SHC values span multiple orders of magnitude, its symmetric log transformation was used as the target ML property.
%with standard ML routines such as 5-fold cross-validation (CV), recursive feature elimination (RFE) and space transformation to symmetric log scale \cite{webber_bi-symmetric_2012}.
The top performing ML model (ridge regression) was then used to make predictions on the candidate dataset of $\sim$ 2000 2D material systems from the cleaned MC2D database. More details on DFT, TB and ML modeling as well as cleaning process for the MC2D database are included in the Methods section.

\begin{figure}[h!]
    \centering
    \includegraphics[width=1\linewidth]{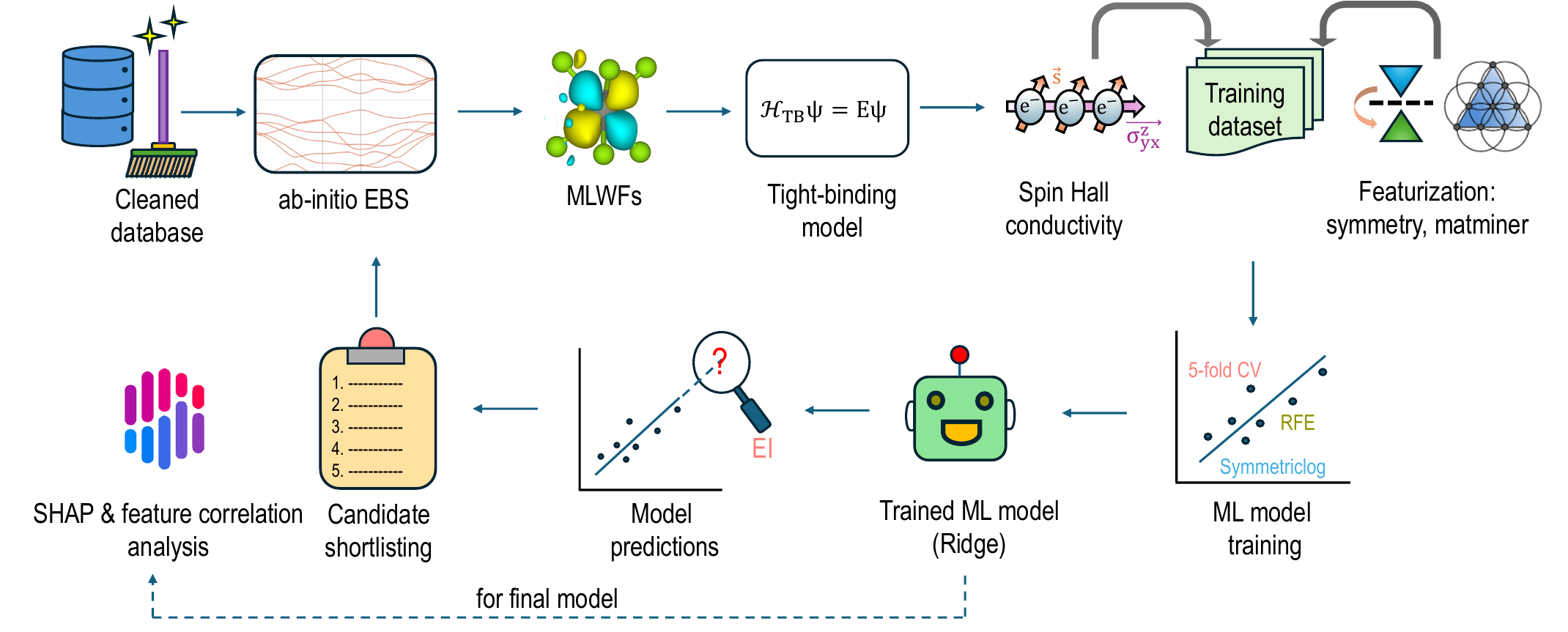}
    \captionsetup{justification=justified}
    \caption{\textbf{Active machine learning framework to identify 2D materials with high SHC.} The workflow primarily consists of DFT (for EBS computation), MLWF (for minimal basis set formation), TB modeling (for computation of Berry curvature and thereby SHC) and ML techniques (for SHC prediction and material selection). First, 2D structures are obtained from MC2D database for which EBS is determined. This is followed by MLWF construction using projection matrices  % $A_{mn}^{k}$ \textcolor{red}{(projection of trial orbital \textit{n} on Bloch band \textit{m} at \textit{k} point of the reciprocal space)}
    and overlap matrices %  $M_{mn}^{k,b}$ \textcolor{red}{(overlap between Bloch bands \textit{m} and \textit{n} at neighboring \textit{k} points connected by the displacement vector \textit{b})}
    generated from EBS using Wannier90 code\cite{mostofi_updated_2014}. The best-fit TB model Hamiltonian is then set up using MLWF basis %\textcolor{red}{and their corresponding hopping parameters $t_{mn}^{j}$ (hopping from orbital \textit{m} to \textit{n} along vector \textit{j})}
    to obtain Berry curvature %$\Omega_{n}^{z}$ \textcolor{red}{(superscript \textit{z} and subscript \textit{n} indicates spin direction and band index respectively)} 
    and SHC. These values are used to create training dataset which also comprises of 26 hand-crafted symmetry and electronic features as well as 132 \textit{matminer} library-based elemental features. The ML models are then trained using 5-fold cross validation (CV), recursive feature elimination and symmetric log scaling of the target property. %After training, the model is tested for predictions on candidate dataset cleaned by handling missing values and inconsistencies (single element compositions, unusual materials with inert elements, etc.).
    The model with lowest CV error is used to screen 5-7 best candidates using expected improvement acquisition strategy for the next round of DFT--SOC+TB computations. The freshly computed SHC data is are added to the training dataset and next active learning loop is initiated.
    %Model delivers \textit{expected improvement} metric for each prediction to help shortlist 5-7 best candidates for further DFT+MLWF+TB computation as a part of active learning loop.
    The final ML model is later subjected to SHAP feature analysis to extract important chemical insights.}
    \label{fig:workflow}
\end{figure}

\par Active learning framework was adopted to efficiently find 2D materials with high SHC. For this, \textit{expected improvement} (EI) acquisition function was adopted which is known to balance the trade-off between exploration (materials with high prediction uncertainty) and exploitation (materials with high predicted SHC value). 
%The objective function to be evaluated in this work is SHC and we have used an acquisition function called \textit{Expected Improvement} (EI) to guide the search for solution under Bayesian optimization methodology.\cite{mockus_bayesian_1975}  This strategy balances the search between unchartered region (exploration) and over-focussing on known region (exploitation); it quantifies EI based on current best value using surrogate model (Ridge) predictions.
Once the ML predictions are available with their corresponding EI values, a few systems (5-7) with high EI values were shortlisted for further SHC computations using the DFT--SOC+TB scheme. Besides EI, the candidates were also selected based on their diversity, i.e., based on number of elements (binary, ternary or quaternary systems) or type of elements (e.g. metal, non-metal). %based on higher EI (also assisted by domain knowledge) for SHC computation (ground truth values) by DFT+MLWF+TB method.
The computed SHC values were again added to the initial training dataset to retrain next version of ML model with better prediction accuracy. This active learning loop were performed for three iterations. Furthermore, to interpret the final trained ML model and the factors that contribute to high SHC in 2D materials, SHapley Additive exPlanations (SHAP) \cite{lundberg_unified_2017} framework was utilized.

\par Figure 2 presents the full set of 2D materials investigated across the three active learning rounds, along with their DFT--SOC+TB computed SHC values. In Round 1 (Figure 2(a)), 24 diverse systems were selected based on domain knowledge and prior literature to capture chemical, structural and property diversity. %variety in type of materials is incorporated based on biased domain knowledge to diversify the training set of the hand-picked 24 systems.
Chemical diversity was introduced by selecting material systems from diverse families of metal halides (4), halonitrides (2), chalcogenides (6), transition metal dichalcogenides (7) and others (5) such as oxychloride, azide, etc.; the numbers in parentheses indicate the count per category. Diversity in terms of various crystal symmetries (12 different space groups, 18 symmorphic, 6 non-symmorphic, 16 centrosymmetric), symmetry elements (inversion, roto-inversion, etc.) as well as polymorphism (two phases of ZrNI) were also considered. Systems spanning low to high SOC and metallic to semiconducting behavior were also included. The t-SNE map (Figure S1) further illustrates the compositional diversity of the Round 1 training data. Figure 2(a) also highlights 2D materials selected in subsequent rounds based on EI values for detailed DFT--SOC+TB computations. These candidates were selected from a large candidate pool of $\sim$2000 2D materials, encompassing diverse range of elements (heavy/light or $s$, $p$, $d$-block), structures from 58 different space groups (SGs), and of compositions ranging from binary to hexanary. Notably, the ML model trained on Round 1 data, which only contains binary materials systems, was successfully used to make predictions for ternary to hexanary materials. Since DFT--SOC+TB computations become exceedingly expensive for pentanary and hexanary systems \cite{tsirkin_high_2021}, we restricted our ML selection to binary, ternary and quaternary systems.
%Figure 2 (a) also shows candidate systems selected (based on their EI values) in the subsequent rounds for DFT+TB computations using the ML model predictions from a large candidate pool of $\sim$2000 2D materials, consisting of a variety of elements (XX) and of systems displaying a large number (58) space groups and multinary classes (binary to hexanary).
%Notably, the ML model has predicted systems (in round 2.1, 2.2 and 3) belonging to the ternary and quaternary classes as well, although the training set is imbalanced (mostly binary and heavy elements based), sparse (24 samples) and high-dimensional (159 features > 24 samples). The model has been subjected to make predictions and evaluate EI on a candidate dataset separated into classes from binary to hexanary for convenience. However, SHC computations are not performed on pentanary and hexanary systems due to complexities (e.g. complicated Fermi surface)\cite{tsirkin_high_2021} as well as expensive nature (due to large number of projections). 

\par Generally, it is difficult to distinguish between Hall conductivity originating from spin or orbital interactions in experimental studies \cite{sala_giant_2022}. However, for the purpose of preliminary screening, we defined total Hall conductivity (THC) as sum of SHC and OHC as a quantitative metric. Accordingly, in our first active learning loop we attempted to build ML models to predict both THC as well as SHC. Round 2.1 and Round 2.2 contain candidates that were selected based on EI values for THC and SHC, respectively, obtained using the same initial training dataset (Round 1) of 24 systems. The actual THC and SHC values for these candidates were subsequently computed from the electronic structure. As can be observed from Figure 2(b), Round 2.2 successfully resulted in identification of candidates with high SHC, much higher than the those in the initial training data. This is evident from the right shift of the orange colored region (Round 2.2) as compared to the black colored region (Round 1). \textcolor{black}{However, this is not the case for THC where green colored region (Round 2.1) is seen to marginally shift up compared to black colored region (Round 1). %Figure S1 and S2 in supplementary information (SI) - 1 shows the poor performance metrics as well as candidate performance of models trained on THC (Round 2.1).
While it is unclear why the performance of the ML model is superior for the case of SHC, a potential cause could be the feature set used in this work.
Thus, for the subsequent rounds we exclusively focused on design of materials with high SHC only.} The SHC data from 36 candidates of Round 1 and 2 (i.e., both Rounds 2.1 and 2.2) was used to train next round of ML model and screen candidates for Round 3. Again, the right shift of blue region (Round 3) compared to orange region (Round 2) indicates the success of the active learning approach to find candidates with higher SHC. The success of the active learning approach to continually find candidates with higher and higher SHC with each subsequent loop is also illustrated using the box-and-whiskers plot in Figure 2(c). This clearly demonstrates the power of combining active ML approaches with traditional computational methods to accelerate discovery of materials. More details on the ML model performance and the identified top candidates is presented next.

\begin{figure}[h!]
    \centering
    \includegraphics[width=1\linewidth]{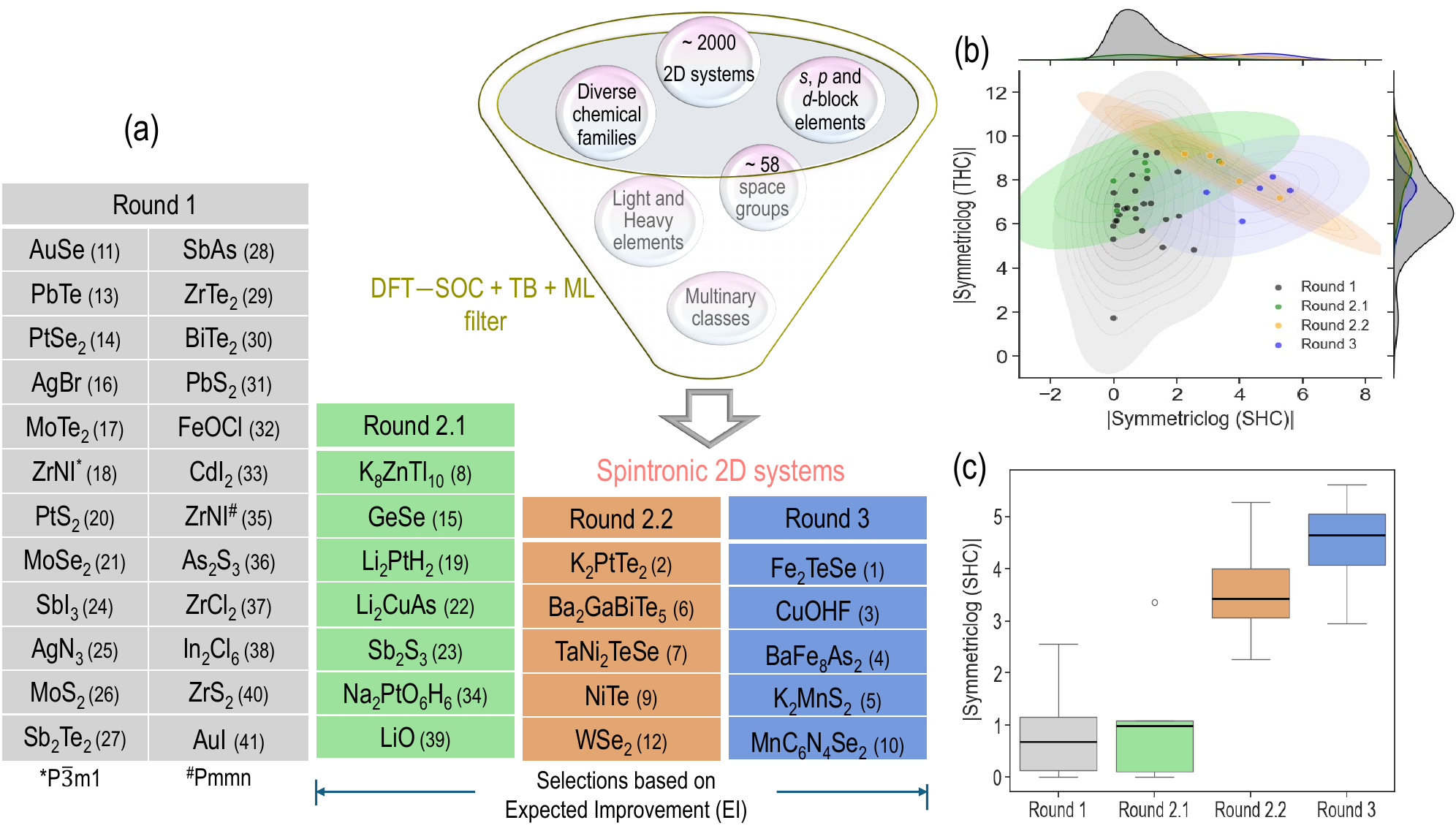}
    \captionsetup{justification=justified}
    \caption{\textbf{Performance of active learning rounds}: (a) 2D systems selected across different rounds for SHC (or THC) computations using active learning approach. While diverse candidates were manually selected in Round 1, EI was used for selection in Rounds 2.1, 2.2 and 3.
    %) for SHC and THC computation by DFT, MLWF and TB studies for active learning strategy with ML.
    The number in parenthesis next to the system name indicates its rank when ordered based on absolute SHC value. (b) Variation in joint kernel density and multi-class scatter plot of THC and SHC with different active learning rounds in absolute symmetric log space. (c) The box-and-whisker plot of modulus of SHC in symmetric log space with horizontal lines in the box showing median, first and third quartiles whereas whiskers (range of the data) are 1.5 times inter-quartile range. The data points beyond the whiskers are outliers marked as circles.}
    \label{fig:result}
\end{figure}  

%median (q\textsubscript{2}), first (q\textsubscript{1}) and third (q\textsubscript{3}) quartiles. Here w\textsubscript{lower} = max(Min data value, q\textsubscript{1} - 1.5 x iqr),  w\textsubscript{upper} = min(Max data value, q\textsubscript{3} + 1.5 x iqr) and iqr = q\textsubscript{3} - q\textsubscript{1}. The datapoints beyond (> q\textsubscript{3} + 1.5 x iqr or < q\textsubscript{1} - 1.5 x iqr) the whiskers are outliers

\begin{figure}[h!]
    \centering
    \includegraphics[width=0.5\linewidth]{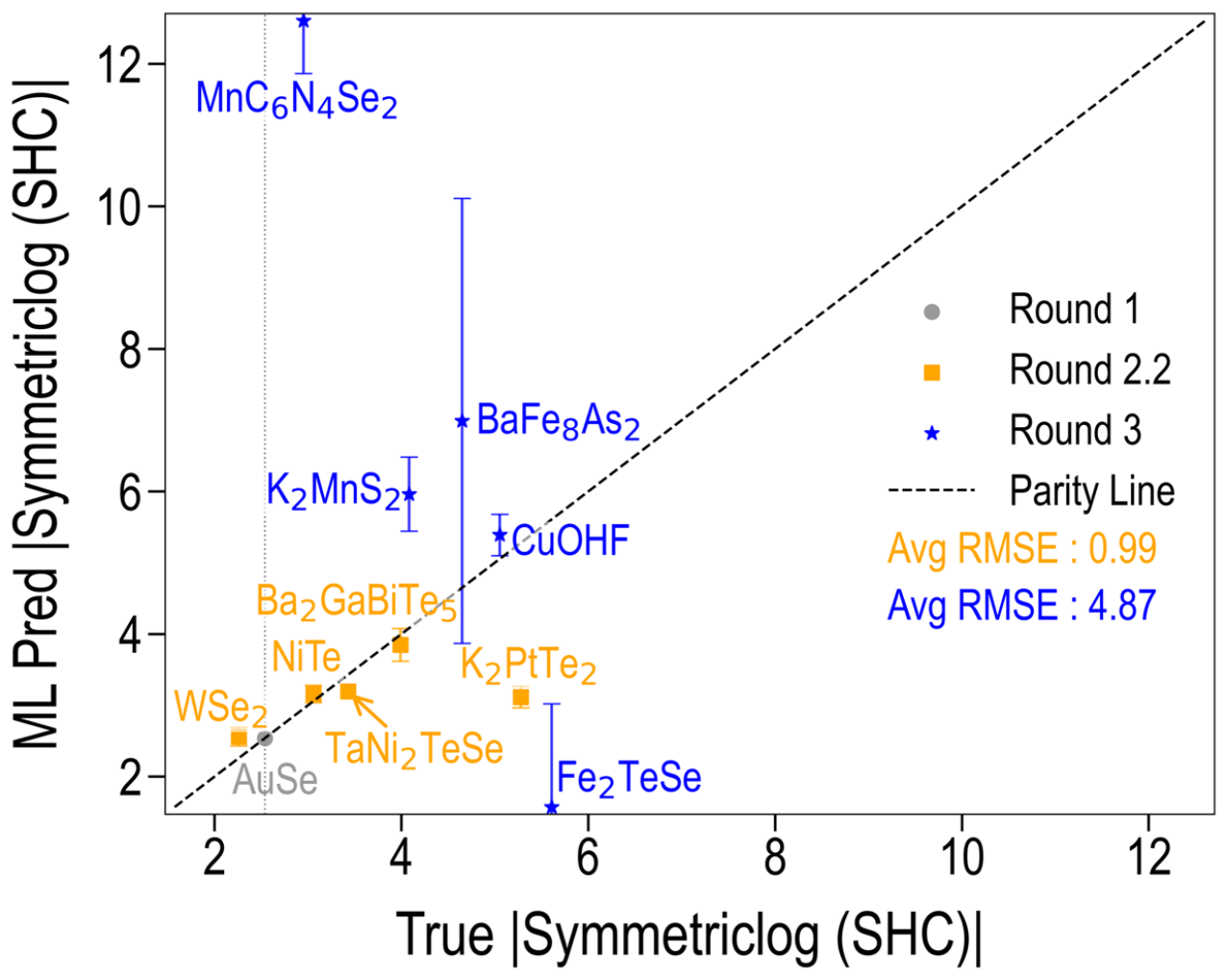}
    \captionsetup{justification=justified}
    \caption{\textbf{Predictive capability of ML Models.} Parity plot comparing the ML predictions against true (computed) SHC values in modulus symmetric log space. Results are shown for candidates screened in Round 2.2 (yellow squares) and Round 3 (blue triangles). The error bars indicate uncertainty ($\pm$1$\sigma$) in model predictions. The diagonal and vertical dashed line  denote the parity line and the SHC value for best candidate (AuSe) from Round 1, respectively. A general trend of discovery of 2D systems with increasing SHC magnitude with each active learning loop can be observed.}
    \label{fig:performance}
\end{figure}

\subsection*{ML model performance}
In each active learning loop, three types of ML models were trained, namely, ridge regression, support vector regression and Gaussian process regression. Ridge regression consistently yielded the lowest 5-fold cross-validation errors across all rounds (see Figure S2 in Supplementary Information-1) and was thus selected for candidate screening in each cycle. Model uncertainty was estimated by training 100 model variants using different random 80\% subsamples of the training data. % Uncertainty in model predictions were obtained by training 100 different variants of a model by randomly sampling (with different seed values) 80 \% of the training data in each round.

%\textcolor{red}{It is to be noted that the primary objective of this work is efficient materials discovery rather than precise property prediction. Thus, validating model performance via a hold-out test set is not the most appropriate strategy for the discovery task and limited training dataset (owing to computational budget and time constraint). Instead,} 
A more stringent and pertinent measure of performance of the ML model is to compare the ML model predictions for screened candidates against their subsequent DFT--SOC+TB computed values. As captured in Figure 3, the ML model generally perform well, with many predictions lying close to the parity line. Two key observations should be noted about this plot. First, it is inherently difficult for the model to obtain good accuracy on such candidates as these are the cases with extreme SHC values and thus are expected to lie in regions where less amount of training data is available. Second, the candidates were screened using EI acquisition function which also involves uncertainty in model prediction. As can be seen, many candidates (Fe$_2$TeSe and BaFe$_8$As$_2$) lying far from the parity line shows high level of uncertainty in model prediction. Nonetheless, %notable discrepancies remain for certain compounds (MnC6N4Se2), indicating limitations in model generalization.
there are also cases (MnC$_6$N$_4$Se$_2$) where the ML model predictions are significantly different than the computed SHC values, highlighting limitations in model generalization. %\textcolor{red}{This is a likely consequence as ML models in discovery tasks are intentionally challenged with candidates lying outside the original data distribution, and it is well established \cite{xue_accelerated_2016, khalak_chemical_2022} that predictive accuracy of ML models for such extreme values is inherently limited. Nevertheless,}
\par More importantly, Figure 3 also captures the progression of candidates with high SHC values identified across different active learning loops. %\textcolor{red}{which is more important and validates the robustness of the trained models. The reliability of models is further supported by low 5-fold CV error (\textcolor{red}{Figure S2} in Supplementary Information-1) and the appearance of physically meaningful descriptors (e.g., atomic number, valence \textit{d} electrons, etc.) in the SHAP analysis in following sub-section. To illustrate,}
In Round 1, AuSe exhibited the highest absolute SHC of 11.7 $\hbar$/e $\Omega$\textsuperscript{-1}. The best performing candidate of Round 2.2, K\textsubscript{2}PtTe\textsubscript{2}, shows an approximately 16-fold increase over the best candidate from Round 1 with SHC of 195.56 $\hbar$/e $\Omega$\textsuperscript{-1}. Further, in Round 3, the best performing candidate Fe\textsubscript{2}TeSe has absolute SHC approximately 1.4 times higher (271.52 $\hbar$/e $\Omega$\textsuperscript{-1}) than that of K\textsubscript{2}PtTe\textsubscript{2}. Overall, the candidate from Round 3 (i.e., Fe\textsubscript{2}TeSe) shows highest SHC among 41 samples and is 23 times higher than the best of Round 1. This consistent improvement across rounds demonstrates the effectiveness of the active learning strategy in discovering 2D materials with better SHC. %Although the SHC values of these top candidates are smaller than conventional 3D heavy metals Pt, Pd, and Au,\cite{guo_ab_2009}.

It is to be noted that the primary objective of this work is efficient materials discovery rather than precise property prediction. Thus, we mostly concentrate on validating if the model iteratively discovers candidates with higher SHC than in the training data set in each round. This is conceptually different from a traditional ML setting wherein the accuracy of the model is validated via a hold-out test set. This process is commonly adopted in materials discovery tasks \cite{xue_accelerated_2016, khalak_chemical_2022} owing to limited training data size and exorbitant cost associated with ground truth generation.

\begin{figure}[h!]
    \centering
    \includegraphics[width=1\linewidth]{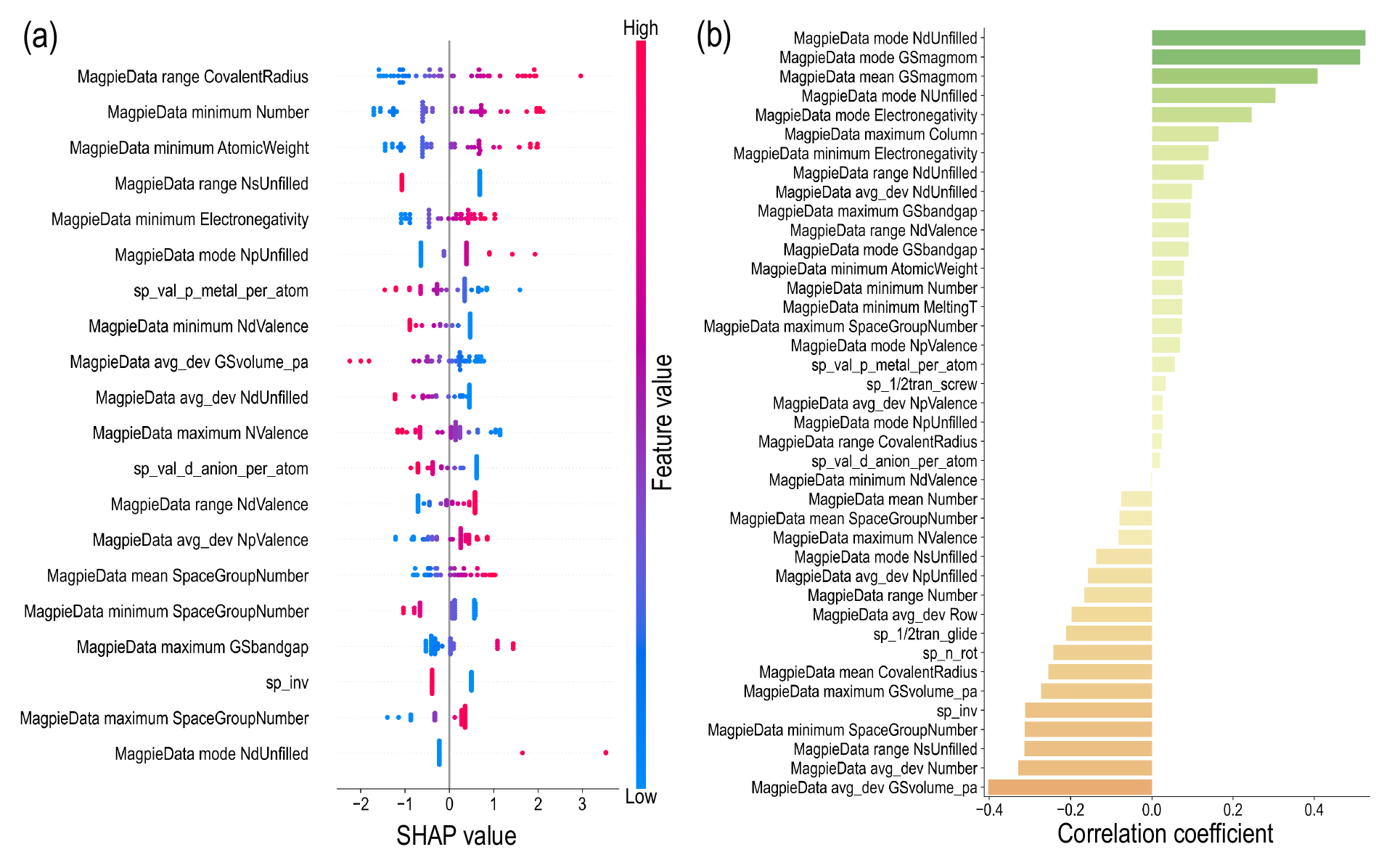}
    \captionsetup{justification=justified}
    \caption{\textbf{Chemical insights from ML model} (a) SHAP summary plot highlighting the most influential features for the final ML model trained on 41 SHC computations (in symlog space). %Each point represents the SHAP value of a feature, with color indicating the original value of feature (red = high, blue = low). Features are ordered by their mean |SHAP| value i.e. their overall impact on the model output.
    (b) The Pearson correlation coefficients between features retained after RFE for the final model and Symlog(SHC). The variation in saturation of color (green and orange) indicates variation in intensity of correlation (positive and negative) respectively.
    %Bars are colored by correlation strength and direction, with green indicating positive and red showing negative correlations. This dual analysis reveals both nonlinear and linear dependencies, offering insights into feature relevance and underlying physical trends.
    }
    \label{fig:analysis}
\end{figure}

\subsection*{Chemical insights from ML models and Data}
To understand the impact of top features on the model predictions, we performed SHAP analysis using the final ML model trained on the full dataset of 41 systems with the results presented in Figure 4(a). A high SHAP value signifies a strong contribution of a given feature to the predicted SHC. Interestingly, both hand-crafted and \textit{Magpie} features were found to be important, indicating contribution to SHC is dominated by a diverse category of features. 
%The magnitude of SHAP value indicates how influential a feature is for prediction of the target property i.e. SHC, whereas negative and positive sign denotes whether it decreases or increases the prediction value.
Notably, \textit{MagpieData range CovalentRadius}, defined as the difference between the maximum and minimum elemental covalent radius in a compound, was found to be the topmost feature. This is consistent with the observation that top ranked SHC candidates comprise of elements from the far right and the far left of the periodic table. We observe that majority of them have heavy elements as constituent atoms which lead to high SOC influence on electronic structure leading to high SHC. Similarly, Magpie features such as minimum atomic number and minimum atomic weight among the constituent elements of the compounds are among the top features. If these numbers are high, it is natural that the compound is made by heavy elements and as said before in such cases the SOC is very high. Further, several features (Magpie and hand-crafted) capturing statistics associated with electronic structure, such as mean, mode or range of unfilled \textit{s}, \textit{p} and \textit{d} orbitals were also found to be important. It is consistent with the understanding that partially filled orbitals are involved in strong covalent interactions which form dispersive bandstructures. When these bands are SOC-split they produce large spin Berry curvature, which is conducive for SHC. The presence or absence of inversion symmetry in the candidate structure was also found to be an important feature. This is in line with the observation that non-centrosymmetric materials show strong Rashba SOC and often result in enhanced SHC. Overall, the convergence of SHAP-identified important features with well-established physical principles %such as the importance of atomic number, electronic configuration, and structural symmetry, 
provides a strong validation of the predictive capacity and physical relevance of the trained ML models.

 \par Similar to SHAP analysis, Pearson correlation plots of features (with SHC in symmetric log space) also shows the important role played by electronic structure of the constituting elements and the crystal symmetry of the 2D materials system (see Figure 4(b)). For instance, statistics (mean, mode, range) associated with unfilled $d$ and $p$ electrons of the elemental systems showed high magnitude of correlation with the computed SHC. Similarly, presence of different symmetry features such as inversion center (\textit{sp\_inv}), different number of rotational axes (\textit{sp\_n\_rot}), a half unit translation in screw axis or glide plane (\textit{sp\_1/2tran\_screw}) displayed negative correlation with SHC, which is understandable since with lowering symmetry the bands tend to be flatter leading to weak quantum transport. Also, lower symmetry can enable more non-zero components of the SHC tensor \cite{roy_unconventional_2022}. Further, in line with the results obtained in this work, a recent theoretical study also found that the absence of inversion symmetry could favor large OHC, which ultimately result in high SHC owing to SOC \cite{sahu_effect_2021}. Other important features that showed large magnitude of correlation with SHC include mean and mode of magnetic moment, mode of electronegativity, and average deviation of ground state volume of constituting elements.

Beyond SHAP and correlation analysis, several other interesting physical trends emerged from the computed SHC data. For instance, all of the top 10 identified 2D systems, except for K\textsubscript{8}ZnTl\textsubscript{10} (ranked 8th), contained at least one metalloid (Ge, As, Sb, or Te) or chalcogen (O, S, Se, or Te). Furthermore, with the exception of NiTe (ranked 9th), all of the top 10 ranked systems were ternary or quaternary, indicating that complex bandstructures may be a better playground for achieving higher SHC. It was also observed that 17 out of the 41 candidates studied in this work were metallic (i.e., zero band gap), and, notably, 9 out of top 10 ranked SHC candidates exhibit metallic character. This is in-line with the observation that conventional and highest SHC bulk materials, e.g., Pt, Pd and Au are also metals \cite{guo_ab_2009}. %Band gap can significantly influence properties that are sensitive to the states near Fermi energy (E\textsubscript{F}) and recent studies also suggest that band gap can significantly alter Berry curvature, thereby impacting SHC \cite{lessnich_spin_2024, hastuti_theoretical_2024}. 
%This is because in metallic systems, the abundance of available electronic states near E\textsubscript{F} increases the likelihood of spin–orbit interactions, which enhances SHC. 
Additionally, the metallic character enables rapid variation in properties such as Hall conductivity with even minor shifts in E\textsubscript{F}, further impacting SHC. This behavior is illustrated in Figure 5, where sharp fluctuations in Hall conductivity near E\textsubscript{F} can be observed for the best candidates identified from each active learning loop, all of which are metallic.

\begin{figure}[!htb]
    \centering
    \includegraphics[width=1\linewidth]{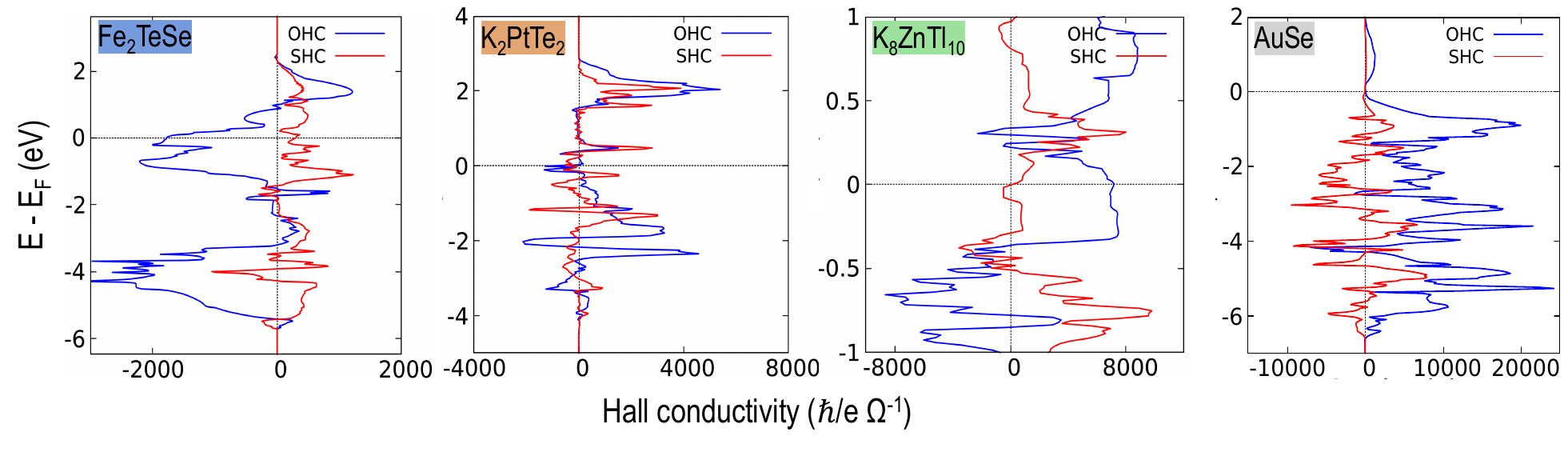}
    \captionsetup{justification=justified}
    \caption{\textbf{Variation of Hall conductivity with energy.} The plot shows energy-dependent variation of orbital (OHC) and spin (SHC) Hall conductivity for best candidate systems from each round computed using TB model \(\mathcal{H}_{\text{TB}}\) and Kubo formalism. The abscissa of point on curves at which zero energy (E-E\textsubscript{F}) horizontal dashed line intersects are taken as OHC and SHC values.}
    \label{fig:shcfermi}
\end{figure}
\par In continuation of previous paragraph, we further elaborate, through Figure 5, the variation of OHC and SHC as a function of change in chemical potential $\mu$ which is measured with respect to E\textsubscript{F}. While OHC and SHC values at E\textsubscript{F} are sufficient to describe the orbital and spin transport, it is also important to consider that $\mu$ is subject to change in real systems based on electron or hole doping, which can arise due to several factors such as interfacial charge transfer, gate voltage tuning, etc. As $\mu$ changes, it alters the relative occupations of the bands and, in turn, affects the net orbital and spin magnetic moments in the Brillouin zone. These moments play an important role in controlling the magnitude of the conductivities. We found that some of the metallic systems indicate a sharp change in conductivities near E\textsubscript{F}, as can be seen in the case of K$_2$PtTe$_2$ (see Figure 5(b)). Such behavior can be excellent in device fabrication, as it points toward a high tunability of the transport properties with a minimal charge doping. In contrast, there are also compounds such as K$_8$ZnTl$_{10}$ that show a much more stable behavior, which can also be useful in generating a constant spin / orbital current that remains unaffected by doping. Furthermore, comparing the orbital and spin transport, we observe that the OHC always dominates over the SHC throughout all occupations. This is expected, as for the non-magnetic compounds, the SHC can only arise from OHC through the spin orbit coupling. However, one must keep in mind that the experimental detection usually involves a ferromagnetic overlayer, where both the orbital and spin currents are passed. While the spin current can directly exert torque, the orbital current must convert to an effective spin current to produce spin-orbit torque\cite{go_orbital_2020}. Although the variation of SHC with the Fermi energy is of the academic interest to examine various possible doping scenarios, in this work, we consider its values at $E_F$ (set to zero) for the ML analysis, as this is typically considered as the standard reference value when reporting the SHC of a material.
%This highly depends on the conversion efficiency of the ferromagnetic material. As a result,in the experiments, it is more often the spin current that produces the SOT.
%for metallic systems, larger number of states are available for electrons to occupy, thereby increasing the possibilities of interaction between spin and orbital degree of freedom, thereby favoring higher SHC. Furthermore, metallic character allows for rapid change in properties such as Hall conductivity with a slight change in E\textsubscript{F} thereby enabling large SHC. This is visualized in Figure 5, where Hall conductivity can be seen to undergo rapid wiggles near E\textsubscript{F} for best candidates from each active learning loop, all of which are metallic. 
\par The dominant presence of \textit{d}-orbitals near E\textsubscript{F} was also found to be an important factor influencing SHC. For example, EBS of 7 out of top 10 candidates, including rank-1 Fe\textsubscript{2}TeSe, is heavily dominated by \textit{d} orbitals of transition elements (Fe, Pt, Cu, Mn, Ta, Zn and Ni) near E\textsubscript{F} (see Figure S3-S4 and Table S1 in Supplementary Information-1 and 2 resp.). Previous studies have shown that transition metal oxides with considerable SOC and dominant \textit{d} orbital character near E\textsubscript{F} exhibit large SHC owing to the SOC enabled mixing of d orbitals 
%which higher degrees of freedom of \textit{d} orbital favor geometrical property $\Omega_{n}^{z}$ of electronic bands
\cite{jadaun_rational_2020}. Evidently, it can be observed that all top-10 ranked candidates except Ba\textsubscript{2}GaBiTe\textsubscript{5} exhibits the combination of metallicity (zero band gap) and dominance of \textit{d}-orbital near E\textsubscript{F}. %Yet another advantage of presence of \textit{d} orbital is their localized character which serves to enhances the electron correlation in a system which is already confined to 2D geometry \cite{meitei_electron_2024}. 
%This can increase SOC and thereby SHE depending on spin Berry curvature and local band occupations \cite{jadaun_rational_2020}. 
Another interesting observation we made in the context of presence of heavy elements. Generally, heavy elements (atomic number $>$ 30) are expected to show high SOC and thus high SHC, but instead we found 7 of the top 10 candidates having dominant \textit{d} orbital contribution from non-heavy elements near E\textsubscript{F} in our dataset. These systems include Fe\textsubscript{2}TeSe (Fe-\textit{d}), CuOHF (Cu-\textit{d}), BaFe\textsubscript{8}As\textsubscript{2} (Fe-\textit{d}), K\textsubscript{2}MnS\textsubscript{2} (Mn-\textit{d}), TaNi\textsubscript{2}TeSe (Ta-\textit{d} and Ni-\textit{d}), NiTe (Ni-\textit{d}) and MnC\textsubscript{6}N\textsubscript{4}Se\textsubscript{2} (Mn-\textit{d}). Among these, it is noteworthy that CuOHF (rank-3) and K\textsubscript{2}MnS\textsubscript{2} (rank-5) does not contain any heavy element at all. On the other hand, there were also candidates that contained heavy elements, e.g., AuI (2 eV band gap), In\textsubscript{2}Cl\textsubscript{6} (3.8 eV), and Na\textsubscript{2}PtO\textsubscript{6}H\textsubscript{6} (3 eV) but showed negligible SHC. This is attributed the fact that in this compounds the orbital Hall effect is estimated to be very small. As a result, the converted SHE is negligible. %This could be due to presence of large band gap or orbitals of heavy elements not dominating the E\textsubscript{F}.

 %For transition metal oxides, it has been theoretically demonstrated that if E\textsubscript{F} locates in an energy window where sub-manifolds of \textit{d} orbital mix under the influence of SOC and crystal field splitting, the material can generate giant SHC.\cite{jadaun_rational_2020} 

 %We observed candidates with at least one type of atomic species of non-zero magnetic moment showing higher SHC. The 6 out of 7 such candidates (\textasciitilde 86\%) secure their spot among top 10 SHC candidates.

Although the absence of inversion symmetry is generally considered favorable for SHC as it creates an intrinsic electric field, from our top 10 candidates we find that it is possible to create large SHC even without breaking inversion symmetry. Similarly, the presence of multiple mirror planes, typically associated with high SHC in bulk materials \cite{zhang_different_2021}, did not consistently correlate with enhanced SHC in our dataset. For instance, while the best candidate Fe\textsubscript{2}TeSe exhibits 4 mirror planes, other systems %(MoS\textsubscript{2}, MoSe\textsubscript{2}, MoTe\textsubscript{2}, WSe\textsubscript{2}, Li\textsubscript{2}CuAs and ZrCl\textsubscript{2})
with 4 mirror planes showed low SHC (< \textasciitilde 10 $\hbar$/e $\Omega$\textsuperscript{-1}). Importantly, a more prominent trend emerged regarding rotoinversion symmetry (i.e., symmetry operation involving rotation followed by inversion). None of the top 10 candidates exhibited rotoinversion symmetry, with the sole exception of NiTe (ranked 9th). Additionally, materials belonging to non-symmorphic space groups, characterized by fractional translation symmetries such as screw axes and glide planes, generally showed low SHC (< 10 $\hbar$/e $\Omega$\textsuperscript{-1}), with K\textsubscript{2}MnS\textsubscript{2} (rank-5, SHC = 58.26 $\hbar$/e $\Omega$\textsuperscript{-1}) being a notable exception.
%Notably, we did find an important trend which is that none of the top 10 candidates displayed rotoinversion symmetry, i.e., presence of both rotational and inversion symmetry together, except NiTe (rank-9). We also found systems from non-symmorphic space group (with fractional translation symmetry such as in screw and glide) exhibit negligible to small SHC  (< 10 $\hbar$/e $\Omega$\textsuperscript{-1}), except for K\textsubscript{2}MnS\textsubscript{2} (rank-2 with SHC of 58.26 $\hbar$/e $\Omega$\textsuperscript{-1}).
It is also interesting to note that several non-symmorphic systems (AuI, %In\textsubscript{2}Cl\textsubscript{6}
ZrNI, FeOCl, AgBr, GeSe) %AuSe 
contain a heavy element, but still exhibit low SHC. It is generally known that lesser symmetry in multilayer structures favors SHC \cite{zhu_large_2022, macneill_control_2017}, but we did not observe such trend based on the 41 2D systems we studied. A particularly illustrative case is ZrNI, which was studied in two space groups, a higher symmetry, centrosymmetric orthorhombic \textit{P$\bar{3}$m1} (rank-18) and low symmetry, non-symmorphic trigonal \textit{Pmmn} (rank-35). Notably, the high-symmetry phase, featuring additional rotational axes, mirror planes, and rotoinversion symmetry, exhibited a higher SHC. 

%\textcolor{red}{Overall, a few suggestive insights can be derived from above trends. The spintronic promise of low SOC systems (2 systems among top-5 do not contain heavy element at all) including 3\textit{d} electron systems (recurrence of Mn, Fe, Ni and Cu in 7 among top-10) can not be ruled out \textit{prima facie}. The impact of specific symmetry elements such as rotoinversion as well as screw axis or glide plane (non-symmorphism) on SHC warrants more attention in future detailed investigations. Most importantly, materials with the combination of metallic nature and dominant \textit{d} character (from heavy or non-heavy elements) near E\textsubscript{F} in absence of roto-inversion symmetry can show greater possibility of demonstrating high spin Hall response. The 7 systems (Fe\textsubscript{2}TeSe, K\textsubscript{2}PtTe\textsubscript{2}, CuOHF, BaFe\textsubscript{8}As\textsubscript{2}, K\textsubscript{2}MnS\textsubscript{2}, TaNi\textsubscript{2}TeSe, and MnC\textsubscript{6}N\textsubscript{4}Se\textsubscript{2}) from top-10 exhibit this combination and 8 out of total such 10 systems (among 41) show relatively better SHC i.e., $>$ 10 $\hbar$/e $\Omega$\textsuperscript{-1}.} 
Contrary to the general expectations of heavy elements (Z > 30) being crucial for high SHC, we found 2 systems, i.e., CuOHF (rank-3) and K\textsubscript{2}MnS\textsubscript{2} (rank-5) (among top 5) that do not contain any heavy elements. Similarly, 7 of the top 10 systems contain contributions from 3\textit{d} transition elements rather than generally expected 5\textit{d} systems. Absence of inversion center or presence of multiple mirror planes is generally considered favorable for high SHC, however, no such trend was observed. Overall, the triple combination of metallic nature, dominant \textit{d} character (from heavy or non-heavy elements) near E\textsubscript{F} and absence of roto-inversion symmetry was found to show high SHC. The 7 systems -- Fe\textsubscript{2}TeSe, K\textsubscript{2}PtTe\textsubscript{2}, CuOHF, BaFe\textsubscript{8}As\textsubscript{2}, K\textsubscript{2}MnS\textsubscript{2}, TaNi\textsubscript{2}TeSe, and MnC\textsubscript{6}N\textsubscript{4}Se\textsubscript{2} -- from top 10 exhibited this combination. Thus, our analysis suggests that there are multiple underlying mechanisms involved in determining the SHC and they are distinct for each class of compounds. Such an observation opens up further scope for fundamental research in this domain.
Another point to be noted is that all of the identified top 10 candidates fall in the category of `easily' or `potentially' exfoliable, as defined in the MC2D work \cite{mounet_two-dimensional_2018}.
Further, we found the rank-1 system Fe\textsubscript{2}TeSe to show dynamic stability (Figure S5 in Supplementary Information-1), and thus top candidates identified in this work are recommended for further detailed computational or experimental investigation. %\textcolor{red}{For the context of synthesizability of promising candidates, it is to noted that the above top-10 candidates fall within the “easily exfoliable” or “potentially exfoliable” categories.\cite{mounet_two-dimensional_2018}} %\textcolor{red}{The rank-1 Fe\textsubscript{2}TeSe showed dynamic stability (\textcolor{red}{Figure S5} in supplementary information-1) and thus top candidates can be subjected for further independent investigations as recent state-of-the-art experimental techniques are capable of synthesize such 2D materials. Although magnetism does not build the necessary criteria for SHE, further scope lies in examining the magnetic aspect of these systems as it can augment the spin transport.}%Rather than looking from symmetry stand-point, this trend is better explained by greater presence of \textit{d} orbital near E\textsubscript{F} and lower band gap in the orthorhombic polymorph as compared to its low symmetric counterpart.

To conclude, we developed the ML algorithm to scan around 2000 2D materials and predict the ones with high SHC. The trained and validation data for the ML algorithm were obtained from DFT calculations. The designed code, which is available on GitHub, is capable of predicting SHC of any other 2D materials. Our analysis emphasizes that the crystal symmetry and SOC are not the only decisive factor for a material to generate spin current via SHE. In addition, other EBS details such as orbital symmetry (\textit{p} or \textit{d}) near E\textsubscript{F}, types of atomic species (metal or non-metal) and composition (binary or multinary), covalent radius or electronegativity of these atomic species, etc., can dictate the phenomenon. It should also be noted that the final ML model uses all of the above properties to make an accurate prediction, and no single dominant property alone can be used to arrive at a good estimate of SHC in 2D systems. %We note that some of above reasoning can not be generalized for all 2D systems as no 1:1 or systematic relationship between SHC and any property is observed.
Further, the active learning strategy employed in this work successfully identified high-performing 2D systems with multifold increase in SHC as compared to initial training data.
  %surpassing identification based on biased human domain knowledge and intuition.
%\textcolor{red}{We note the model’s algorithmic limitation in capturing OHC, likely due to constraints in the features employed, and recommend constrained interpretation of the derived chemical insights given the limited dataset size.}
We anticipate that present work shall help both the spintronic and the broader material informatics community. The set of high SHC candidates identified in this work can be further studied. At the same time, this work presents a successful example of ML guided accelerated discovery wherein the ML models are able to successfully find superior systems with desired properties from a large candidate pool.
  %studied in detail using with the catalytic recipe for accelerated material discovery along with insights for their tentative design to some extent. 

  %And, candidates with higher order of rotational axis in rotoinversion symmetry populate in bottom 75\% of 41 systems. 

\section*{METHODS}
\subsection*{Computational details}

\textbf{DFT and TB workflow}. The DFT computations were performed using Quantum ESPRESSO \cite{giannozzi_quantum_2009}, and FINDSYM \cite{stokes_findsym_2005} was used for symmetry analysis. The exchange and correlation behavior of electron-electron interactions were approximated under the generalized gradient approximation in the Perdew-Burke-Ernzerhof (GGA-PBE) \cite{perdew_generalized_1996} method. Whereas for nucleus-electrons interactions, the norm-conserving pseudopotentials \cite{hamann_optimized_2013} (from PseudoDojo library \cite{van_setten_pseudodojo_2018}) were used owing to their reputation of exact eigenvalues and node-less eigenfunctions to better compute the Berry curvature. The 2D crystal structures, imported from the Materials Cloud Two-Dimensional Structure Database (MC2D)\cite{mounet_two-dimensional_2018, campi_expansion_2023}, were modeled using a slab approach with sufficient vacuum along \textit{z} direction (> 12~\AA) to avoid periodic interactions. Although MC2D database already contains relaxed structures, in the initial round a few structures were re-relaxed. Since only negligible changes were observed, the re-relaxation step was omitted in the subsequent rounds. %These systems were relaxed prior to SHC computations, if needed.
A sufficiently high and plane-wave basis cut-off for wavefunction expansion (in similar range of PseudoDojo library recommended values)  and charge-density was used along with relevant total energy convergence cut-off.
%\textcolor{red}{The workflow used to compute SHC and OHC is identical for all 41 systems as mentioned in following steps.} 
\\
\indent (i) We first perform DFT computations to obtain EBS without SOC. For this, we performed self-consistent field (SCF) calculations using an automatic \textit{k}-mesh with varied \textit{k}-spacing in a range of 0.03-0.17~\AA \textsuperscript{-1}.  This is followed by non-SCF calculations to analyze partial density of states (Figure S3 in Supplementary Information-1 and Table S1 in
Supplementary Information-2) to choose appropriate projections and compute wavefunctions on full uniform \textit{k}-mesh with varied \textit{k}-spacing 0.02-0.10~\AA \textsuperscript{-1} as inputs to construction of MLWFs using Wannier90 \cite{mostofi_updated_2014} package. MLWFs are constructed in the absence of SOC to avoid mix-up of wavefunctions and the error in the eigenstates. High-quality MLWFs were ensured by achieving spread parameters up to 10\textsuperscript{-12}~\AA\textsuperscript{2}. After post-processing accurately fit MLWFs (Figure S4 in Supplementary Information-1 and Table S1 in
Supplementary Information-2), the hopping parameters were obtained for step-iii. \\
\indent (ii) Separate EBS calculations including SOC
effects are performed to estimate SOC strengths as extra inputs to step-iii.\\
\indent (iii) Kubo formalism for orbital and spin transport:
In this sub-section, we discuss the computation of the orbital and spin transport properties. First,  a numerical TB model was constructed based on the hopping parameters obtained from step-i. Keeping the accuracy of the eigenstates in mind, the spin orbit coupling was initially turned off. Later, SOC from step-ii was introduced as an additional term to the Hamiltonian, and the SOC parameters were tuned to match DFT results. We introduced the SOC as an extra parameter to the Hamiltonian and later the band structure was fit with the DFT EBS with SOC (Figure S6 in  Supplementary Information-1), by tuning the atomistic SOC parameter $\lambda$. This fitting validates not just the accuracy of the eigenvalues (bands), but also the eigenstates. This method is commonly used in the literature in the context of theoretical tight-binding models \cite{go_intrinsic_2018, bhowal_intrinsic_2020, sahu_emergence_2024}. Table S1 of Supplementary Information-1 lists $\lambda$ obtained for all the materials.
The matrix elements of the Hamiltonian are given by,
\begin{equation}
    H_{mn} = \sum_j t_{mn}^j e^{i\vec{k}\cdot d_j} +\lambda \vec{L}\cdot\vec{S},
\end{equation}
where \textit{m} and \textit{n} are the basis indices, \textit{j} runs over all the neighbors and $t_{mn}^j$ corresponds to the hopping parameter. In the second term, the parameter $\lambda$ represents the SOC strength. The eigenvalues and eigenstates of this Hamiltonian were further used to calculate the orbital (OHC) and spin Hall conductivity (SHC) on a dense \textit{k}-grid of $800\times800\times1$. The Kubo formula \cite{sinova_universal_2004} for the conductivities can be expressed as,
%As mentioned earlier, the MLWFs were taken from without-SOC DFT computations whereas with-SOC DFT computations provided SOC-based parameters to set-up the TB model Hamiltonian. 

\begin{equation}\label{OHC} 
  \sigma^{\gamma,\rm orb/spin}_{\alpha \beta}   =  -\frac{e} { N_k V_c} \sum_{ \vec  k} \Omega^{\gamma,\rm orb/spin}_{\alpha \beta} ({\vec  k}),
\end{equation}
where $\Omega^{\gamma,\rm orb/spin}_{\alpha \beta} ({\vec  k})
= \sum_n^{\rm occ} \Omega^{\gamma,\rm orb/spin}_{n,\alpha \beta} ({\vec  k})$ is the sum over occupied bands $n$ at a specific momemtum point $\vec k$. Here, $\alpha,\beta$, and $\gamma$ can be $x,y$ or $z$ such that $\alpha$ is the direction of the orbital/spin current, $\beta$ is the direction of the applied electric field, while the orbital and spin angular momentum points towards $\gamma$, and $V_c$ corresponds to the volume of the unit cell. Importantly, for 2D systems the volume term is replaced by the area of the unit cell. Therefore, the unit of SHC for 2D systems is usually expressed in $\hbar$/e $\Omega$\textsuperscript{-1}.    \cite{matusalem_quantization_2019}
The orbital Berry curvature for the Bloch state $n\vec k$ is computed using the Kubo formula,
\begin{equation} \label{obc}         
 \Omega^{\gamma,\rm orb}_{n,\alpha\beta} ({\vec  k}) = 2 \hbar   \sum_{n^\prime \neq n} \frac {{\rm Im}[ \langle u_{n{\vec  k}} | \mathcal{J}^{\gamma,\rm orb}_\alpha | u_{n^\prime{\vec  k}} \rangle  
\langle u_{n^\prime{\vec  k}} | v_\beta | u_{n{\vec  k}} \rangle]} 
{(\varepsilon_{n^\prime \vec k}-\varepsilon_{n  \vec k} )^2},
\end{equation}
where $v_{\alpha} =  \frac{1}{\hbar} \frac{\partial {\cal H} }{ \partial k_\alpha}$ is the velocity operator as defined earlier, and the orbital current operator is $\mathcal{J}^{\gamma,\rm orb}_\alpha = \frac{1}{2} \{v_\alpha, L_\gamma \}$,  
with $L_\gamma$ being the orbital angular momentum. For the spin Berry curvature, the orbital current operator is replaced by the spin current operator $\mathcal{J}^{\gamma,\rm spin}_\alpha = \frac{1}{2} \{v_\alpha, S_\gamma \}$ in Eq. (\ref{obc}). In the present case, for the 2D systems, $\sigma_{yx}^z$ is the only conventional component of the OHC and SHC tensor that survives, which is used throughout this work. It is also important to note that both the orbital and spin Hall effects can produce spin orbit torque in the experiments, which is measured by a non-vanishing spin Hall angle. 
To validate our computational approach, we compared the calculated SHC values of 2D TMDs with a  prior work. \cite{bhowal_intrinsic_2020} As summarized in Table S2 of Supplementary Information-1, our results show excellent agreement with the literature.

\textbf{Machine learning workflow}.
%In following section, we outline the ML pipeline developed to identify materials with high SHC. Due to the limited availability of data, we employed an active learning loop that iteratively trained the models, selected the most promising candidates, and retrained with updated data. Multiple iterations of this process resulted in a refined dataset and a model capable of predicting materials with high SHC.
%The initial dataset of 24 randomly selected 2D materials was sourced from the MC2D database. MC2D is a widely used resource in computational materials science domain that comprises of nearly 2000 2D material compositions and their corresponding crystal and electronic structure properties. Before proceeding with ML model development, the MC2D dataset was refined by removing entries with missing band gap, containing only single element, or elements for which pseudo-potentials were not available. This resulted in curated final clean candidate pool of 1947 2D material systems. Two category of input features were used to train the ML models: Magpie features, generated via the \textit{matminer} library, that capture elemental properties like electronegativity, covalent radius, volume, valence electron counts across s, p, d, and f orbitals, and their statistical variants such as mean, mode, range, etc.; and hand-crafted features meant to capture crystal symmetry of the 2D system such as rotation, inversion, or roto-inversion symmetry. A total of 158 features, consisting of 134 Magpie and 24 hand-crafted features, were used for initial model development.
An initial dataset comprising 24 randomly selected 2D materials was extracted from the MC2D database, a widely utilized resource in computational materials science that includes approximately 2,000 2D material compositions along with their associated crystallographic and electronic structure data. Prior to ML model development, the dataset was systematically curated by excluding entries that lacked band gap information, consist only of a single element, or included elements for which norm-conserving pseudopotentials were unavailable. This pre-processing yielded a final, cleaned candidate pool of 1,947 2D material systems.

Two categories of input features were employed to train the ML models. The first consisted of Magpie features, generated using the \textit{matminer} library, which encode elemental properties such as electronegativity, covalent radius, atomic volume, and valence electron counts across the 
\textit{s}, \textit{p}, and \textit{d} orbitals, along with their statistical descriptors (e.g., mean, mode, range). The second category comprised of hand-crafted features designed to capture crystallographic symmetry of the 2D material system, including the presence of rotational, inversion, or roto-inversion symmetries. In total, 158 features, 26 hand-crafted (Table S3 in Supplementary Information-1) and 132 Magpie-derived, were used in the initial phase of model training. %The detailed list of features with the description of their acronyms is provided in Table S1 and S2 in Supplementary Information-1.

%The SHC and THC values were calculated for each material using the previously detailed method. A candidate set, distinct from the 24 training materials, was fetched from the MC2D database. This set was refined by removing entries with missing band gap data, single-element compositions, or absence of pseudopotentials, etc. yielding a curated final candidate pool of \textcolor{red}{1947} materials. This curated candidate set was specifically used to identify materials with high SHC values as described further.

%Model training utilized two feature sets: Magpie features, generated via the \textit{matminer} library, which accounts of atomistic properties like electronegativity and valence electron counts across s, p, d, and f orbitals; and hand-crafted crystal symmetry features derived from crystallographic information, which capture structural attributes such as rotations, inversion, etc. The model was trained on a total of 158 features: 134 Magpie features from \textit{matminer} and 24 symmetry based features.

%Furthermore, the initial dataset was found to be biased towards binary composition materials. To address this, we ensured that candidate selection included a diverse range of compositions, thus promoting comprehensive representation of materials possibility.

\par To ensure generalizability with a limited dataset size, relatively simple ML models, including ridge regression, Gaussian process regression (GPR), and support vector regression (SVR), were employed for model development. Given that the SHC (and THC) values span several orders of magnitude, a symmetric logarithmic (Symlog) transformation \cite{webber_bi-symmetric_2012} given by the equation \( y = \operatorname{sign}(x) \cdot \ln\left(1 + \lvert x/C \rvert \right) \)
 was applied (with \textit{C} = 1) to the target property prior to training. To reduce feature dimensionality and improve model generalization, a recursive feature elimination (RFE) strategy was used. Features resulting in highest 5-fold cross-validation (CV) error were recursively removed until the CV error began to increase. The feature set (Table S4 in Supplementary Information-1) corresponding to the minimum CV error was retained for subsequent model training. Hyperparameter optimization was performed individually for each model type to further enhance performance. Model evaluation was conducted using 5-fold CV, with average root mean squared error (RMSE) and average coefficient of determination (R²) serving as performance metrics. Among the tested models, ridge regression with feature normalization and RFE consistently yielded best performance across all active learning loops. The CV R² score (and RMSE in symlog units) across active learning rounds 1, 2.2, 3 and 4 was 0.94 (1.58), 0.98 (0.11), 0.94 (0.39) and 0.94 (0.46), respectively, demonstrating the consistently high predictive accuracy of the trained models. The parity plots comparing the performance of all ML models across different active learning loops are included in Figure S2 in Supplementary Information-1.

This optimized ridge regression model was then used to predict SHC for materials remaining in the candidate pool, excluding those cases which are part of the training dataset. To estimate prediction uncertainty, an ensemble of 100 models was generated by randomly sampling 80\% of the training data for each variant. The expected improvement (EI) acquisition function, which balances exploration and exploitation, was employed to select the most promising candidates for subsequent DFT--SOC + TB modeling. Besides EI, diversity of the candidates in terms of type (light or heavy, alkali,  alkaline earth or transition metals, metalloid, chalcogens or halogens) or number (binary, ternary or quaternary) of elements was also considered to screen next round of candidates.

\section*{DATA AVAILABILITY}
The SHC data generated in this work can be accessed at the \href{https://github.com/MI-LAB-IITM/Active-Learning-Guided-Computational-Discovery-of-2D-Materials-with-Large-Spin-Hall-Conductivity.git}{GitHub Repository}. The visualisation of training data using t-SNE map, list of features (before and after RFE), comparative performance of different ML models across active learning rounds, SOC parameters for all 41 systems, and a representative example of DFT+SOC band fitting with TB (including SOC), and comparison of calculated SHC values of few TMDs with literature is given in Supplementary Information-1. The MLWF fitting of \textit{ab-initio} EBS for top 6 ranked systems and their PDOS are also provided in Supplementary Information-1, whereas those for rank-7 to rank-41 materials are provided in Supplementary Information-2.

\section*{CODE AVAILABILITY}

The trained ML model, \textcolor{black}{along with the complete dataset} of this work, is publicly available at \href{https://github.com/MI-LAB-IITM/Active-Learning-Guided-Computational-Discovery-of-2D-Materials-with-Large-Spin-Hall-Conductivity.git}{GitHub Repository}.

%\highlightrefs{\cite{xue_accelerated_2016, khalak_chemical_2022}}
%\redrefs{refkey47}
%\redrefs{refkey58}
%\redrefs{refkey59}

%\bibliography{refs}

\bibliography{references}

@article{matusalem_quantization_2019,
	title = {Quantization of spin {Hall} conductivity in two-dimensional topological insulators versus symmetry and spin-orbit interaction},
	volume = {100},
	url = {https://link.aps.org/doi/10.1103/PhysRevB.100.245430},
	doi = {10.1103/PhysRevB.100.245430},
	number = {24},
	urldate = {2026-06-11},
	journal = {Physical Review B},
	publisher = {American Physical Society},
	author = {Matusalem, Filipe and Marques, Marcelo and Teles, Lara K. and Matthes, Lars and Furthmüller, Jürgen and Bechstedt, Friedhelm},
	month = dec,
	year = {2019},
	pages = {245430},
}

@article{bhowal_intrinsic_2020,
	title = {Intrinsic orbital and spin {Hall} effects in monolayer transition metal dichalcogenides},
	volume = {102},
	url = {https://link.aps.org/doi/10.1103/PhysRevB.102.035409},
	doi = {10.1103/PhysRevB.102.035409},
	number = {3},
	urldate = {2026-06-11},
	journal = {Physical Review B},
	publisher = {American Physical Society},
	author = {Bhowal, Sayantika and Satpathy, S.},
	month = jul,
	year = {2020},
	pages = {035409},
}

@article{sahu_emergence_2024,
	title = {Emergence of giant orbital {Hall} and tunable spin {Hall} effects in centrosymmetric transition metal dichalcogenides},
	volume = {110},
	url = {https://link.aps.org/doi/10.1103/PhysRevB.110.054403},
	doi = {10.1103/PhysRevB.110.054403},
	number = {5},
	journal = {Phys. Rev. B},
	publisher = {American Physical Society},
	author = {Sahu, Pratik and Bidika, Jatin Kumar and Biswal, Bubunu and Satpathy, S. and Nanda, B. R. K.},
	month = aug,
	year = {2024},
	pages = {054403},
}

@article{go_intrinsic_2018,
	title = {Intrinsic {Spin} and {Orbital} {Hall} {Effects} from {Orbital} {Texture}},
	volume = {121},
	url = {https://link.aps.org/doi/10.1103/PhysRevLett.121.086602},
	doi = {10.1103/PhysRevLett.121.086602},
	number = {8},
	journal = {Phys. Rev. Lett.},
	publisher = {American Physical Society},
	author = {Go, Dongwook and Jo, Daegeun and Kim, Changyoung and Lee, Hyun-Woo},
	month = aug,
	year = {2018},
	pages = {086602},
}

@article{khalak_chemical_2022,
	title = {Chemical {Space} {Exploration} with {Active} {Learning} and {Alchemical} {Free} {Energies}},
	volume = {18},
	issn = {1549-9618},
	url = {https://doi.org/10.1021/acs.jctc.2c00752},
	doi = {10.1021/acs.jctc.2c00752},
	number = {10},
	journal = {Journal of Chemical Theory and Computation},
	publisher = {American Chemical Society},
	author = {Khalak, Yuriy and Tresadern, Gary and Hahn, David F. and de Groot, Bert L. and Gapsys, Vytautas},
	month = oct,
	year = {2022},
	pages = {6259--6270},
}

@article{xue_accelerated_2016,
	title = {Accelerated search for materials with targeted properties by adaptive design},
	volume = {7},
	issn = {2041-1723},
	url = {https://doi.org/10.1038/ncomms11241},
	doi = {10.1038/ncomms11241},
	number = {1},
	journal = {Nature Communications},
	author = {Xue, Dezhen and Balachandran, Prasanna V. and Hogden, John and Theiler, James and Xue, Deqing and Lookman, Turab},
	month = apr,
	year = {2016},
	pages = {11241},
}

@article{van_setten_pseudodojo_2018,
	title = {The {PseudoDojo}: {Training} and grading a 85 element optimized norm-conserving pseudopotential table ({MMC} {S2}. {Notebook} providing the analysis of the test results of the {ONCVPSP} {PBE} v0.4 {PseudoDojo} tables.)},
	volume = {226},
	issn = {0010-4655},
	shorttitle = {The {PseudoDojo}},
	url = {https://www.sciencedirect.com/science/article/pii/S0010465518300250},
	doi = {10.1016/j.cpc.2018.01.012},
	urldate = {2025-10-08},
	journal = {Computer Physics Communications},
	author = {van Setten, M. J. and Giantomassi, M. and Bousquet, E. and Verstraete, M. J. and Hamann, D. R. and Gonze, X. and Rignanese, G. -M.},
	month = may,
	year = {2018},
	pages = {39--54},
}

@article{zhou_high-throughput_2025,
	title = {High-throughput calculations of spin {Hall} conductivity in non-magnetic {2D} materials},
	volume = {9},
	issn = {2397-7132},
	url = {https://doi.org/10.1038/s41699-025-00562-4},
	doi = {10.1038/s41699-025-00562-4},
	number = {1},
	journal = {npj 2D Materials and Applications},
	author = {Zhou, Jiaqi and Poncé, Samuel and Charlier, Jean-Christophe},
	month = may,
	year = {2025},
	pages = {39},
}

@article{nanda_electronic_2012,
	title = {Electronic structure of the substitutional vacancy in graphene: density-functional and {Green}'s function studies},
	volume = {14},
	url = {https://dx.doi.org/10.1088/1367-2630/14/8/083004},
	doi = {10.1088/1367-2630/14/8/083004},
	number = {8},
	journal = {New Journal of Physics},
	publisher = {IOP Publishing},
	author = {Nanda, B R K and Sherafati, M and Popović, Z S and Satpathy, S},
	month = aug,
	year = {2012},
	pages = {083004},
}

@article{padmanabhan_intertwined_2016,
	title = {Intertwined lattice deformation and magnetism in monovacancy graphene},
	volume = {93},
	url = {https://link.aps.org/doi/10.1103/PhysRevB.93.165403},
	doi = {10.1103/PhysRevB.93.165403},
	number = {16},
	journal = {Phys. Rev. B},
	publisher = {American Physical Society},
	author = {Padmanabhan, Haricharan and Nanda, B. R. K.},
	month = apr,
	year = {2016},
	pages = {165403},
}

@article{campi_expansion_2023,
	title = {Expansion of the {Materials} {Cloud} {2D} {Database}},
	volume = {17},
	issn = {1936-0851},
	url = {https://doi.org/10.1021/acsnano.2c11510},
	doi = {10.1021/acsnano.2c11510},
	number = {12},
	urldate = {2025-01-13},
	journal = {ACS Nano},
	publisher = {American Chemical Society. (MC2D website/database accessed on 10th October 2023)},
	author = {Campi, Davide and Mounet, Nicolas and Gibertini, Marco and Pizzi, Giovanni and Marzari, Nicola},
	month = jun,
	year = {2023},
	pages = {11268--11278},
}

@article{yang_proximity_2013,
	title = {Proximity {Effects} {Induced} in {Graphene} by {Magnetic} {Insulators}: {First}-{Principles} {Calculations} on {Spin} {Filtering} and {Exchange}-{Splitting} {Gaps}},
	volume = {110},
	url = {https://link.aps.org/doi/10.1103/PhysRevLett.110.046603},
	doi = {10.1103/PhysRevLett.110.046603},
	number = {4},
	journal = {Phys. Rev. Lett.},
	publisher = {American Physical Society},
	author = {Yang, H. X. and Hallal, A. and Terrade, D. and Waintal, X. and Roche, S. and Chshiev, M.},
	month = jan,
	year = {2013},
	pages = {046603},
}

@article{cervenka_room-temperature_2009,
	title = {Room-temperature ferromagnetism in graphite driven by two-dimensional networks of point defects},
	volume = {5},
	issn = {1745-2481},
	url = {https://doi.org/10.1038/nphys1399},
	doi = {10.1038/nphys1399},
	number = {11},
	journal = {Nature Physics},
	author = {Červenka, J. and Katsnelson, M. I. and Flipse, C. F. J.},
	month = nov,
	year = {2009},
	pages = {840--844},
}

@article{hong_room-temperature_2012,
	title = {Room-temperature {Magnetic} {Ordering} in {Functionalized} {Graphene}},
	volume = {2},
	issn = {2045-2322},
	url = {https://doi.org/10.1038/srep00624},
	doi = {10.1038/srep00624},
	number = {1},
	journal = {Scientific Reports},
	author = {Hong, Jeongmin and Bekyarova, Elena and Liang, Ping and de Heer, Walt A. and Haddon, Robert C. and Khizroev, Sakhrat},
	month = sep,
	year = {2012},
	pages = {624},
}

@article{nair_dual_2013,
	title = {Dual origin of defect magnetism in graphene and its reversible switching by molecular doping},
	volume = {4},
	issn = {2041-1723},
	url = {https://doi.org/10.1038/ncomms3010},
	doi = {10.1038/ncomms3010},
	number = {1},
	journal = {Nature Communications},
	author = {Nair, R.R. and Tsai, I.-L. and Sepioni, M. and Lehtinen, O. and Keinonen, J. and Krasheninnikov, A.V. and Castro Neto, A.H. and Katsnelson, M.I. and Geim, A.K. and Grigorieva, I.V.},
	month = jun,
	year = {2013},
	pages = {2010},
}

@article{yazyev_defect-induced_2007,
	title = {Defect-induced magnetism in graphene},
	volume = {75},
	url = {https://link.aps.org/doi/10.1103/PhysRevB.75.125408},
	doi = {10.1103/PhysRevB.75.125408},
	number = {12},
	journal = {Phys. Rev. B},
	publisher = {American Physical Society},
	author = {Yazyev, Oleg V. and Helm, Lothar},
	month = mar,
	year = {2007},
	pages = {125408},
}

@article{haugen_spin_2008,
	title = {Spin transport in proximity-induced ferromagnetic graphene},
	volume = {77},
	url = {https://link.aps.org/doi/10.1103/PhysRevB.77.115406},
	doi = {10.1103/PhysRevB.77.115406},
	number = {11},
	journal = {Phys. Rev. B},
	publisher = {American Physical Society},
	author = {Haugen, Håvard and Huertas-Hernando, Daniel and Brataas, Arne},
	month = mar,
	year = {2008},
	pages = {115406},
}

@article{roy_unconventional_2022,
	title = {Unconventional spin {Hall} effects in nonmagnetic solids},
	volume = {6},
	url = {https://link.aps.org/doi/10.1103/PhysRevMaterials.6.045004},
	doi = {10.1103/PhysRevMaterials.6.045004},
	number = {4},
	journal = {Phys. Rev. Mater.},
	publisher = {American Physical Society},
	author = {Roy, Arunesh and Guimarães, Marcos H. D. and Sławiıfmmode {\textbackslash}acuten{\textbackslash}else ń{\textbackslash}fiska, Jagoda},
	month = apr,
	year = {2022},
	pages = {045004},
}

@article{go_orbital_2020,
	title = {Orbital torque: {Torque} generation by orbital current injection},
	volume = {2},
	url = {https://link.aps.org/doi/10.1103/PhysRevResearch.2.013177},
	doi = {10.1103/PhysRevResearch.2.013177},
	number = {1},
	journal = {Phys. Rev. Res.},
	publisher = {American Physical Society},
	author = {Go, Dongwook and Lee, Hyun-Woo},
	month = feb,
	year = {2020},
	pages = {013177},
}

@article{mudgal_magnetic-proximity-induced_2023,
	title = {Magnetic-{Proximity}-{Induced} {Efficient} {Charge}-to-{Spin} {Conversion} in {Large}-{Area} {PtSe2}/{Ni80Fe20} {Heterostructures}},
	volume = {23},
	issn = {1530-6984},
	url = {https://doi.org/10.1021/acs.nanolett.3c04060},
	doi = {10.1021/acs.nanolett.3c04060},
	number = {24},
	journal = {Nano Letters},
	publisher = {American Chemical Society},
	author = {Mudgal, Richa and Jakhar, Alka and Gupta, Pankhuri and Yadav, Ram Singh and Biswal, Bubunu and Sahu, Pratik and Bangar, Himanshu and Kumar, Akash and Chowdhury, Niru and Satpati, Biswarup and Kumar Nanda, Birabar Ranjit and Satpathy, Sashi and Das, Samaresh and Muduli, Pranaba Kishor},
	month = dec,
	year = {2023},
	pages = {11925--11931},
}

@article{wunderlich_experimental_2005,
	title = {Experimental {Observation} of the {Spin}-{Hall} {Effect} in a {Two}-{Dimensional} {Spin}-{Orbit} {Coupled} {Semiconductor} {System}},
	volume = {94},
	url = {https://link.aps.org/doi/10.1103/PhysRevLett.94.047204},
	doi = {10.1103/PhysRevLett.94.047204},
	number = {4},
	journal = {Phys. Rev. Lett.},
	publisher = {American Physical Society},
	author = {Wunderlich, J. and Kaestner, B. and Sinova, J. and Jungwirth, T.},
	month = feb,
	year = {2005},
	pages = {047204},
}

@article{ahn_2d_2020,
	title = {{2D} materials for spintronic devices},
	volume = {4},
	issn = {2397-7132},
	url = {https://doi.org/10.1038/s41699-020-0152-0},
	doi = {10.1038/s41699-020-0152-0},
	number = {1},
	journal = {npj 2D Materials and Applications},
	author = {Ahn, Ethan C.},
	month = jun,
	year = {2020},
	pages = {17},
}

@article{song_coexistence_2020,
	title = {Coexistence of large conventional and planar spin {Hall} effect with long spin diffusion length in a low-symmetry semimetal at room temperature},
	volume = {19},
	issn = {1476-4660},
	url = {https://doi.org/10.1038/s41563-019-0600-4},
	doi = {10.1038/s41563-019-0600-4},
	number = {3},
	journal = {Nature Materials},
	author = {Song, Peng and Hsu, Chuang-Han and Vignale, Giovanni and Zhao, Meng and Liu, Jiawei and Deng, Yujun and Fu, Wei and Liu, Yanpeng and Zhang, Yuanbo and Lin, Hsin and Pereira, Vitor M. and Loh, Kian Ping},
	month = mar,
	year = {2020},
	pages = {292--298},
}

@article{sinova_universal_2004,
	title = {Universal {Intrinsic} {Spin} {Hall} {Effect}},
	volume = {92},
	url = {https://link.aps.org/doi/10.1103/PhysRevLett.92.126603},
	doi = {10.1103/PhysRevLett.92.126603},
	number = {12},
	journal = {Phys. Rev. Lett.},
	publisher = {American Physical Society},
	author = {Sinova, Jairo and Culcer, Dimitrie and Niu, Q. and Sinitsyn, N. A. and Jungwirth, T. and MacDonald, A. H.},
	month = mar,
	year = {2004},
	pages = {126603},
}

@article{li_first-principles_2016,
	title = {First-principles design of spintronics materials},
	volume = {3},
	issn = {2095-5138},
	url = {https://doi.org/10.1093/nsr/nww026},
	doi = {10.1093/nsr/nww026},
	number = {3},
	journal = {National Science Review},
	author = {Li, Xingxing and Yang, Jinlong},
	month = apr,
	year = {2016},
	note = {\_eprint: https://academic.oup.com/nsr/article-pdf/3/3/365/31566317/nww026.pdf},
	pages = {365--381},
}

@article{sinova_spin_2015,
	title = {Spin {Hall} effects},
	volume = {87},
	url = {https://link.aps.org/doi/10.1103/RevModPhys.87.1213},
	doi = {10.1103/RevModPhys.87.1213},
	number = {4},
	journal = {Rev. Mod. Phys.},
	publisher = {American Physical Society},
	author = {Sinova, Jairo and Valenzuela, Sergio O. and Wunderlich, J. and Back, C. H. and Jungwirth, T.},
	month = oct,
	year = {2015},
	pages = {1213--1260},
}

@article{macneill_control_2017,
	title = {Control of spin–orbit torques through crystal symmetry in {WTe2}/ferromagnet bilayers},
	volume = {13},
	issn = {1745-2481},
	url = {https://doi.org/10.1038/nphys3933},
	doi = {10.1038/nphys3933},
	number = {3},
	journal = {Nature Physics},
	author = {MacNeill, D. and Stiehl, G. M. and Guimaraes, M. H. D. and Buhrman, R. A. and Park, J. and Ralph, D. C.},
	month = mar,
	year = {2017},
	pages = {300--305},
}

@article{zhu_large_2022,
	title = {Large spin hall conductivity in low-symmetry semiconductor {ZrSe3}},
	volume = {918},
	issn = {0925-8388},
	url = {https://www.sciencedirect.com/science/article/pii/S0925838822019703},
	doi = {https://doi.org/10.1016/j.jallcom.2022.165579},
	journal = {Journal of Alloys and Compounds},
	author = {Zhu, Yonghui and Chen, Qian and Wu, Haijing and Liang, Jian and Tian, Mingming and Jiang, Wei and Wang, Jiachen and Li, Rongxin and Li, Shangkun and Huang, Zhaocong and Kou, Zhaoxia and Lv, Weiming and Zhang, Baoshun and Zeng, Zhongming and Zhai, Ya},
	year = {2022},
	pages = {165579},
}

@article{sala_giant_2022,
	title = {Giant orbital {Hall} effect and orbital-to-spin conversion in 3d, 5d, and 4f metallic heterostructures},
	volume = {4},
	url = {https://link.aps.org/doi/10.1103/PhysRevResearch.4.033037},
	doi = {10.1103/PhysRevResearch.4.033037},
	number = {3},
	journal = {Phys. Rev. Res.},
	publisher = {American Physical Society},
	author = {Sala, Giacomo and Gambardella, Pietro},
	month = jul,
	year = {2022},
	pages = {033037},
}

@article{meitei_electron_2024,
	title = {Electron {Correlation} in {2D} {Periodic} {Systems} from {Periodic} {Bootstrap} {Embedding}},
	volume = {15},
	number = {48},
	journal = {The Journal of Physical Chemistry Letters},
	publisher = {ACS Publications},
	author = {Meitei, Oinam Romesh and Van Voorhis, Troy},
	year = {2024},
	pages = {11992--12000},
}

@article{jadaun_rational_2020,
	title = {Rational design principles for giant spin {Hall} effect in 5d-transition metal oxides},
	volume = {117},
	number = {22},
	journal = {Proceedings of the National Academy of Sciences},
	publisher = {National Academy of Sciences},
	author = {Jadaun, Priyamvada and Register, Leonard F and Banerjee, Sanjay K},
	year = {2020},
	pages = {11878--11886},
}

@article{guo_ab_2009,
	title = {Ab initio calculation of intrinsic spin {Hall} conductivity of {Pd} and {Au}},
	volume = {105},
	issn = {0021-8979},
	url = {https://doi.org/10.1063/1.3054362},
	doi = {10.1063/1.3054362},
	number = {7},
	urldate = {2025-05-06},
	journal = {Journal of Applied Physics},
	author = {Guo, G. Y.},
	month = jan,
	year = {2009},
	pages = {07C701},
}

@article{guo_intrinsic_2008,
	title = {Intrinsic {Spin} {Hall} {Effect} in {Platinum}: {First}-{Principles} {Calculations}},
	volume = {100},
	url = {https://link.aps.org/doi/10.1103/PhysRevLett.100.096401},
	doi = {10.1103/PhysRevLett.100.096401},
	number = {9},
	journal = {Phys. Rev. Lett.},
	publisher = {American Physical Society},
	author = {Guo, G. Y. and Murakami, S. and Chen, T.-W. and Nagaosa, N.},
	month = mar,
	year = {2008},
	pages = {096401},
}

@article{sagasta_tuning_2016,
	title = {Tuning the spin {Hall} effect of {Pt} from the moderately dirty to the superclean regime},
	volume = {94},
	url = {https://link.aps.org/doi/10.1103/PhysRevB.94.060412},
	doi = {10.1103/PhysRevB.94.060412},
	number = {6},
	journal = {Phys. Rev. B},
	publisher = {American Physical Society},
	author = {Sagasta, Edurne and Omori, Yasutomo and Isasa, Miren and Gradhand, Martin and Hueso, Luis E. and Niimi, Yasuhiro and Otani, YoshiChika and Casanova, Fèlix},
	month = aug,
	year = {2016},
	pages = {060412},
}

@article{xu_high-throughput_2024,
	title = {High-throughput calculation of large spin {Hall} conductivity in heavy-metal-based antiperovskite compounds},
	journal = {Materials Genome Engineering Advances},
	publisher = {Wiley Online Library},
	author = {Xu, Xiong and Lv, JX and Wang, Y and Li, Min and Wang, Zhe and Wang, Hui},
	year = {2024},
	pages = {e69},
}

@article{ji_spin_2022,
	title = {Spin {Hall} conductivity and anomalous {Hall} conductivity in full {Heusler} compounds},
	volume = {24},
	url = {https://dx.doi.org/10.1088/1367-2630/ac696c},
	doi = {10.1088/1367-2630/ac696c},
	number = {5},
	journal = {New Journal of Physics},
	publisher = {IOP Publishing},
	author = {Ji, Yimin and Zhang, Wenxu and Zhang, Hongbin and Zhang, Wanli},
	month = may,
	year = {2022},
	pages = {053027},
}

@article{zhang_different_2021,
	title = {Different types of spin currents in the comprehensive materials database of nonmagnetic spin {Hall} effect},
	volume = {7},
	issn = {2057-3960},
	url = {https://doi.org/10.1038/s41524-021-00635-0},
	doi = {10.1038/s41524-021-00635-0},
	number = {1},
	journal = {npj Computational Materials},
	author = {Zhang, Yang and Xu, Qiunan and Koepernik, Klaus and Rezaev, Roman and Janson, Oleg and Železný, Jakub and Jungwirth, Tomáš and Felser, Claudia and van den Brink, Jeroen and Sun, Yan},
	month = oct,
	year = {2021},
	pages = {167},
}

@inproceedings{lundberg_unified_2017,
	title = {A {Unified} {Approach} to {Interpreting} {Model} {Predictions}},
	volume = {30},
	url = {https://proceedings.neurips.cc/paper_files/paper/2017/file/8a20a8621978632d76c43dfd28b67767-Paper.pdf},
	booktitle = {Advances in {Neural} {Information} {Processing} {Systems}},
	publisher = {Curran Associates, Inc.},
	author = {Lundberg, Scott M and Lee, Su-In},
	editor = {Guyon, I. and Luxburg, U. Von and Bengio, S. and Wallach, H. and Fergus, R. and Vishwanathan, S. and Garnett, R.},
	year = {2017},
}

@article{webber_bi-symmetric_2012,
	title = {A bi-symmetric log transformation for wide-range data},
	volume = {24},
	url = {https://dx.doi.org/10.1088/0957-0233/24/2/027001},
	doi = {10.1088/0957-0233/24/2/027001},
	number = {2},
	journal = {Measurement Science and Technology},
	publisher = {IOP Publishing},
	author = {Webber, J Beau W},
	month = dec,
	year = {2012},
	pages = {027001},
}

@article{ward_matminer_2018,
	title = {Matminer: {An} open source toolkit for materials data mining},
	volume = {152},
	issn = {0927-0256},
	url = {https://www.sciencedirect.com/science/article/pii/S0927025618303252},
	doi = {https://doi.org/10.1016/j.commatsci.2018.05.018},
	journal = {Computational Materials Science},
	author = {Ward, Logan and Dunn, Alexander and Faghaninia, Alireza and Zimmermann, Nils E. R. and Bajaj, Saurabh and Wang, Qi and Montoya, Joseph and Chen, Jiming and Bystrom, Kyle and Dylla, Maxwell and Chard, Kyle and Asta, Mark and Persson, Kristin A. and Snyder, G. Jeffrey and Foster, Ian and Jain, Anubhav},
	year = {2018},
	pages = {60--69},
}

@article{ren_hybrid_2023,
	title = {Hybrid spin {Hall} nano-oscillators based on ferromagnetic metal/ferrimagnetic insulator heterostructures},
	volume = {14},
	copyright = {2023 The Author(s)},
	issn = {2041-1723},
	url = {https://www.nature.com/articles/s41467-023-37028-4},
	doi = {10.1038/s41467-023-37028-4},
	language = {en},
	number = {1},
	urldate = {2025-04-30},
	journal = {Nature Communications},
	publisher = {Nature Publishing Group},
	author = {Ren, Haowen and Zheng, Xin Yu and Channa, Sanyum and Wu, Guanzhong and O’Mahoney, Daisy A. and Suzuki, Yuri and Kent, Andrew D.},
	month = mar,
	year = {2023},
	pages = {1406},
}

@article{sahu_effect_2021,
	title = {Effect of the inversion symmetry breaking on the orbital {Hall} effect: {A} model study},
	volume = {103},
	url = {https://link.aps.org/doi/10.1103/PhysRevB.103.085113},
	doi = {10.1103/PhysRevB.103.085113},
	number = {8},
	journal = {Phys. Rev. B},
	publisher = {American Physical Society},
	author = {Sahu, Pratik and Bhowal, Sayantika and Satpathy, S.},
	month = feb,
	year = {2021},
	pages = {085113},
}

@article{garrity_database_2021,
	title = {Database of {Wannier} tight-binding {Hamiltonians} using high-throughput density functional theory},
	volume = {8},
	issn = {2052-4463},
	url = {https://doi.org/10.1038/s41597-021-00885-z},
	doi = {10.1038/s41597-021-00885-z},
	number = {1},
	journal = {Scientific Data},
	author = {Garrity, Kevin F. and Choudhary, Kamal},
	month = apr,
	year = {2021},
	pages = {106},
}

@article{vitale_automated_2020,
	title = {Automated high-throughput {Wannierisation}},
	volume = {6},
	issn = {2057-3960},
	url = {https://doi.org/10.1038/s41524-020-0312-y},
	doi = {10.1038/s41524-020-0312-y},
	number = {1},
	journal = {npj Computational Materials},
	author = {Vitale, Valerio and Pizzi, Giovanni and Marrazzo, Antimo and Yates, Jonathan R. and Marzari, Nicola and Mostofi, Arash A.},
	month = jun,
	year = {2020},
	pages = {66},
}

@article{gresch_automated_2018,
	title = {Automated construction of symmetrized {Wannier}-like tight-binding models from ab initio calculations},
	volume = {2},
	url = {https://link.aps.org/doi/10.1103/PhysRevMaterials.2.103805},
	doi = {10.1103/PhysRevMaterials.2.103805},
	number = {10},
	journal = {Phys. Rev. Mater.},
	publisher = {American Physical Society},
	author = {Gresch, Dominik and Wu, QuanSheng and Winkler, Georg W. and Häuselmann, Rico and Troyer, Matthias and Soluyanov, Alexey A.},
	month = oct,
	year = {2018},
	pages = {103805},
}

@article{qiao_calculation_2018,
	title = {Calculation of intrinsic spin {Hall} conductivity by {Wannier} interpolation},
	volume = {98},
	url = {https://link.aps.org/doi/10.1103/PhysRevB.98.214402},
	doi = {10.1103/PhysRevB.98.214402},
	number = {21},
	journal = {Phys. Rev. B},
	publisher = {American Physical Society},
	author = {Qiao, Junfeng and Zhou, Jiaqi and Yuan, Zhe and Zhao, Weisheng},
	month = dec,
	year = {2018},
	pages = {214402},
}

@article{soumyanarayanan_emergent_2016,
	title = {Emergent phenomena induced by spin–orbit coupling at surfaces and interfaces},
	volume = {539},
	issn = {1476-4687},
	url = {https://doi.org/10.1038/nature19820},
	doi = {10.1038/nature19820},
	number = {7630},
	journal = {Nature},
	author = {Soumyanarayanan, Anjan and Reyren, Nicolas and Fert, Albert and Panagopoulos, Christos},
	month = nov,
	year = {2016},
	pages = {509--517},
}

@article{kong_path_2019,
	title = {Path towards graphene commercialization from lab to market},
	volume = {14},
	issn = {1748-3395},
	url = {https://doi.org/10.1038/s41565-019-0555-2},
	doi = {10.1038/s41565-019-0555-2},
	number = {10},
	journal = {Nature Nanotechnology},
	author = {Kong, Wei and Kum, Hyun and Bae, Sang-Hoon and Shim, Jaewoo and Kim, Hyunseok and Kong, Lingping and Meng, Yuan and Wang, Kejia and Kim, Chansoo and Kim, Jeehwan},
	month = oct,
	year = {2019},
	pages = {927--938},
}

@article{shao_strong_2016,
	title = {Strong {Rashba}-{Edelstein} {Effect}-{Induced} {Spin}–{Orbit} {Torques} in {Monolayer} {Transition} {Metal} {Dichalcogenide}/{Ferromagnet} {Bilayers}},
	volume = {16},
	issn = {1530-6984},
	url = {https://doi.org/10.1021/acs.nanolett.6b03300},
	doi = {10.1021/acs.nanolett.6b03300},
	number = {12},
	journal = {Nano Letters},
	publisher = {American Chemical Society},
	author = {Shao, Qiming and Yu, Guoqiang and Lan, Yann-Wen and Shi, Yumeng and Li, Ming-Yang and Zheng, Cheng and Zhu, Xiaodan and Li, Lain-Jong and Amiri, Pedram Khalili and Wang, Kang L.},
	month = dec,
	year = {2016},
	pages = {7514--7520},
}

@article{chernyshov_evidence_2009,
	title = {Evidence for reversible control of magnetization in a ferromagnetic material by means of spin–orbit magnetic field},
	volume = {5},
	issn = {1745-2481},
	url = {https://doi.org/10.1038/nphys1362},
	doi = {10.1038/nphys1362},
	number = {9},
	journal = {Nature Physics},
	author = {Chernyshov, Alexandr and Overby, Mason and Liu, Xinyu and Furdyna, Jacek K. and Lyanda-Geller, Yuli and Rokhinson, Leonid P.},
	month = sep,
	year = {2009},
	pages = {656--659},
}

@article{dai_interfacial_2024,
	title = {Interfacial magnetic spin {Hall} effect in van der {Waals} {Fe3GeTe2}/{MoTe2} heterostructure},
	volume = {15},
	issn = {2041-1723},
	url = {https://doi.org/10.1038/s41467-024-45318-8},
	doi = {10.1038/s41467-024-45318-8},
	number = {1},
	journal = {Nature Communications},
	author = {Dai, Yudi and Xiong, Junlin and Ge, Yanfeng and Cheng, Bin and Wang, Lizheng and Wang, Pengfei and Liu, Zenglin and Yan, Shengnan and Zhang, Cuiwei and Xu, Xianghan and Shi, Youguo and Cheong, Sang-Wook and Xiao, Cong and Yang, Shengyuan A. and Liang, Shi-Jun and Miao, Feng},
	month = feb,
	year = {2024},
	pages = {1129},
}

@article{pai_spin_2012,
	title = {Spin transfer torque devices utilizing the giant spin {Hall} effect of tungsten},
	volume = {101},
	issn = {0003-6951},
	url = {https://doi.org/10.1063/1.4753947},
	doi = {10.1063/1.4753947},
	number = {12},
	urldate = {2025-03-22},
	journal = {Applied Physics Letters},
	author = {Pai, Chi-Feng and Liu, Luqiao and Li, Y. and Tseng, H. W. and Ralph, D. C. and Buhrman, R. A.},
	month = sep,
	year = {2012},
	pages = {122404},
}

@article{kimura_room-temperature_2007,
	title = {Room-{Temperature} {Reversible} {Spin} {Hall} {Effect}},
	volume = {98},
	url = {https://link.aps.org/doi/10.1103/PhysRevLett.98.156601},
	doi = {10.1103/PhysRevLett.98.156601},
	number = {15},
	journal = {Phys. Rev. Lett.},
	publisher = {American Physical Society},
	author = {Kimura, T. and Otani, Y. and Sato, T. and Takahashi, S. and Maekawa, S.},
	month = apr,
	year = {2007},
	pages = {156601},
}

@article{xu_high_2020,
	title = {High {Spin} {Hall} {Conductivity} in {Large}-{Area} {Type}-{II} {Dirac} {Semimetal} {PtTe2}},
	volume = {32},
	url = {https://advanced.onlinelibrary.wiley.com/doi/abs/10.1002/adma.202000513},
	doi = {https://doi.org/10.1002/adma.202000513},
	number = {17},
	journal = {Advanced Materials},
	author = {Xu, Hongjun and Wei, Jinwu and Zhou, Hengan and Feng, Jiafeng and Xu, Teng and Du, Haifeng and He, Congli and Huang, Yuan and Zhang, Junwei and Liu, Yizhou and Wu, Han-Chun and Guo, Chenyang and Wang, Xiao and Guang, Yao and Wei, Hongxiang and Peng, Yong and Jiang, Wanjun and Yu, Guoqiang and Han, Xiufeng},
	year = {2020},
	note = {\_eprint: https://advanced.onlinelibrary.wiley.com/doi/pdf/10.1002/adma.202000513},
	pages = {2000513},
}

@article{tsirkin_high_2021,
	title = {High performance {Wannier} interpolation of {Berry} curvature and related quantities with {WannierBerri} code},
	volume = {7},
	copyright = {2021 The Author(s)},
	issn = {2057-3960},
	url = {https://www.nature.com/articles/s41524-021-00498-5},
	doi = {10.1038/s41524-021-00498-5},
	language = {en},
	number = {1},
	urldate = {2025-03-16},
	journal = {npj Computational Materials},
	publisher = {Nature Publishing Group},
	author = {Tsirkin, Stepan S.},
	month = feb,
	year = {2021},
	pages = {1--9},
}

@article{han_graphene_2014,
	title = {Graphene spintronics},
	volume = {9},
	copyright = {2014 Springer Nature Limited},
	issn = {1748-3395},
	url = {https://www.nature.com/articles/nnano.2014.214},
	doi = {10.1038/nnano.2014.214},
	language = {en},
	number = {10},
	urldate = {2025-03-16},
	journal = {Nature Nanotechnology},
	publisher = {Nature Publishing Group},
	author = {Han, Wei and Kawakami, Roland K. and Gmitra, Martin and Fabian, Jaroslav},
	month = oct,
	year = {2014},
	pages = {794--807},
}

@article{mostofi_updated_2014,
	title = {An updated version of wannier90: {A} tool for obtaining maximally-localised {Wannier} functions},
	volume = {185},
	issn = {0010-4655},
	shorttitle = {An updated version of wannier90},
	url = {https://www.sciencedirect.com/science/article/pii/S001046551400157X},
	doi = {10.1016/j.cpc.2014.05.003},
	number = {8},
	urldate = {2025-03-05},
	journal = {Computer Physics Communications},
	author = {Mostofi, Arash A. and Yates, Jonathan R. and Pizzi, Giovanni and Lee, Young-Su and Souza, Ivo and Vanderbilt, David and Marzari, Nicola},
	month = aug,
	year = {2014},
	pages = {2309--2310},
}

@article{giannozzi_quantum_2009,
	title = {{QUANTUM} {ESPRESSO}: a modular and open-source software project for quantum simulations of materials},
	volume = {21},
	issn = {0953-8984},
	shorttitle = {{QUANTUM} {ESPRESSO}},
	url = {https://dx.doi.org/10.1088/0953-8984/21/39/395502},
	doi = {10.1088/0953-8984/21/39/395502},
	language = {en},
	number = {39},
	urldate = {2025-03-05},
	journal = {Journal of Physics: Condensed Matter},
	author = {Giannozzi, Paolo and Baroni, Stefano and Bonini, Nicola and Calandra, Matteo and Car, Roberto and Cavazzoni, Carlo and Ceresoli, Davide and Chiarotti, Guido L and Cococcioni, Matteo and Dabo, Ismaila and Dal Corso, Andrea and de Gironcoli, Stefano and Fabris, Stefano and Fratesi, Guido and Gebauer, Ralph and Gerstmann, Uwe and Gougoussis, Christos and Kokalj, Anton and Lazzeri, Michele and Martin-Samos, Layla and Marzari, Nicola and Mauri, Francesco and Mazzarello, Riccardo and Paolini, Stefano and Pasquarello, Alfredo and Paulatto, Lorenzo and Sbraccia, Carlo and Scandolo, Sandro and Sclauzero, Gabriele and Seitsonen, Ari P and Smogunov, Alexander and Umari, Paolo and Wentzcovitch, Renata M},
	month = sep,
	year = {2009},
	pages = {395502},
}

@article{mounet_two-dimensional_2018,
	title = {Two-dimensional materials from high-throughput computational exfoliation of experimentally known compounds},
	volume = {13},
	copyright = {2018 © The Author (s) 2017, under exclusive licence to Macmillan Publishers Limited, part of Springer Nature},
	issn = {1748-3395},
	url = {https://www.nature.com/articles/s41565-017-0035-5},
	doi = {10.1038/s41565-017-0035-5},
	language = {en},
	number = {3},
	urldate = {2025-03-05},
	journal = {Nature Nanotechnology},
	publisher = {Nature Publishing Group},
	author = {Mounet, Nicolas and Gibertini, Marco and Schwaller, Philippe and Campi, Davide and Merkys, Andrius and Marrazzo, Antimo and Sohier, Thibault and Castelli, Ivano Eligio and Cepellotti, Andrea and Pizzi, Giovanni and Marzari, Nicola},
	month = mar,
	year = {2018},
	pages = {246--252},
}

@article{hamann_optimized_2013,
	title = {Optimized norm-conserving {Vanderbilt} pseudopotentials},
	volume = {88},
	url = {https://link.aps.org/doi/10.1103/PhysRevB.88.085117},
	doi = {10.1103/PhysRevB.88.085117},
	number = {8},
	urldate = {2025-01-13},
	journal = {Physical Review B},
	publisher = {American Physical Society},
	author = {Hamann, D. R.},
	month = aug,
	year = {2013},
	pages = {085117},
}

@article{perdew_generalized_1996,
	title = {Generalized {Gradient} {Approximation} {Made} {Simple}},
	volume = {77},
	url = {https://link.aps.org/doi/10.1103/PhysRevLett.77.3865},
	doi = {10.1103/PhysRevLett.77.3865},
	number = {18},
	urldate = {2024-05-13},
	journal = {Physical Review Letters},
	publisher = {American Physical Society},
	author = {Perdew, John P. and Burke, Kieron and Ernzerhof, Matthias},
	month = oct,
	year = {1996},
	pages = {3865--3868},
}

@article{stokes_findsym_2005,
	title = {{FINDSYM}: program for identifying the space-group symmetry of a crystal},
	volume = {38},
	issn = {0021-8898},
	shorttitle = {{FINDSYM}},
	url = {//scripts.iucr.org/cgi-bin/paper?zm5027},
	doi = {10.1107/S0021889804031528},
	language = {en},
	number = {1},
	urldate = {2024-05-13},
	journal = {Journal of Applied Crystallography},
	publisher = {International Union of Crystallography},
	author = {Stokes, H. T. and Hatch, D. M.},
	month = feb,
	year = {2005},
	pages = {237--238},
}

\section*{Acknowledgements}
Author A.J.K. gratefully acknowledges financial support previously received from the Center for Atomistic Modelling and Materials Design (CAMMD), IIT Madras and currently from the Anusandhan National Research Foundation (ANRF / erstwhile SERB) \textit{National Post-Doctoral Fellowship} (NPDF) scheme (File No. PDF/2023/000036). We gratefully acknowledge the computing resources provided on \textit{Improv}, \textit{Bebop} and \textit{Swing}, a high-performance computing cluster operated by the Laboratory Computing Resource Center (LCRC) at Argonne National Laboratory (ANL), USA. Authors A.J.K. and R.B. acknowledge the part of this work is performed at \textit{Carbon} cluster (proposal \#83855) at Center for Nanoscale Materials, a U.S. Department of Energy Office of Science User Facility, was supported by the U.S. DOE, Office of Basic Energy Sciences, under Contract No. DE-AC02-06CH11357. R.B. acknowledges ANRF for the funding (File No. SRG/2023/001055). Computational resources from Robert Bosch Centre for Data Science and AI, IIT Madras, are also acknowledged.

\section*{Author contributions statement}
A.J.K.: DFT Computation and MLWF Construction, ML Data Curation, Visualization, Analysis, Writing. \\
S.N.: ML Investigation, ML Data Curation, Visualization, Analysis, Writing. \\
P.S.: TB Investigation and SHC Computation, Validation, Analysis, Writing. \\
H.C.: ML Investigation. \\
B.R.K.N.: Conceptualization, Resources, Funding Acquisition,  Supervision, Validation. \\
R.B.: Conceptualization, Resources, Funding Acquisition, Supervision, Visualization, Validation.  \\
All authors reviewed and edited the manuscript.

\end{document}

\maketitle % This generates the title block

\begin{figure}[h]
    \centering
    \includegraphics[width=0.7\linewidth]{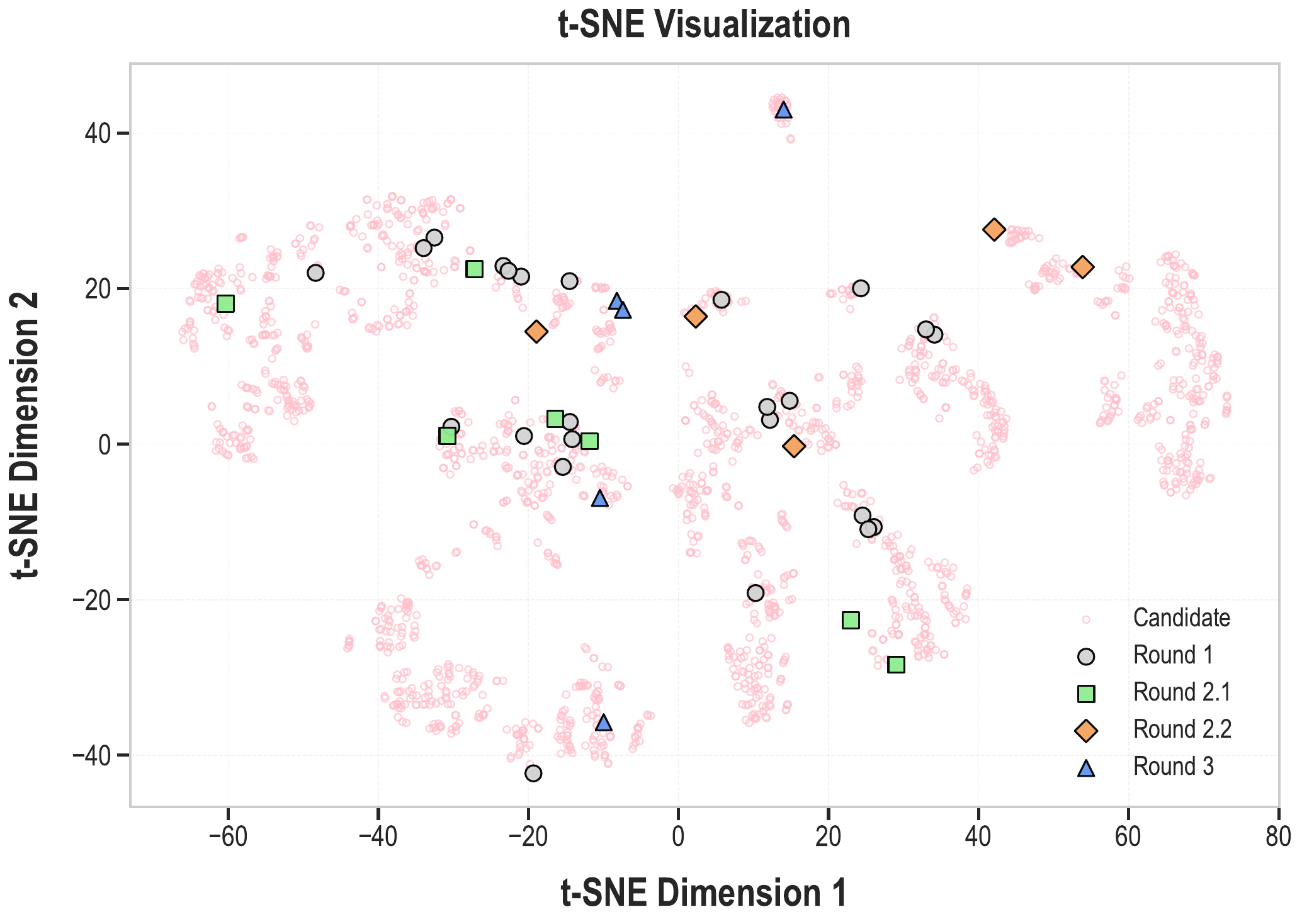}
    \caption{\textcolor{black}{\textbf{The t-SNE map} showing distribution of various material systems selected during different rounds of active learning loop. The Round 1 materials (gray points) are well spread across the feature space, indicating good initial training diversity.}}
    \label{fig:tsne}
\end{figure}

\begin{figure}[h]
    \centering
    \includegraphics[width=\textwidth]{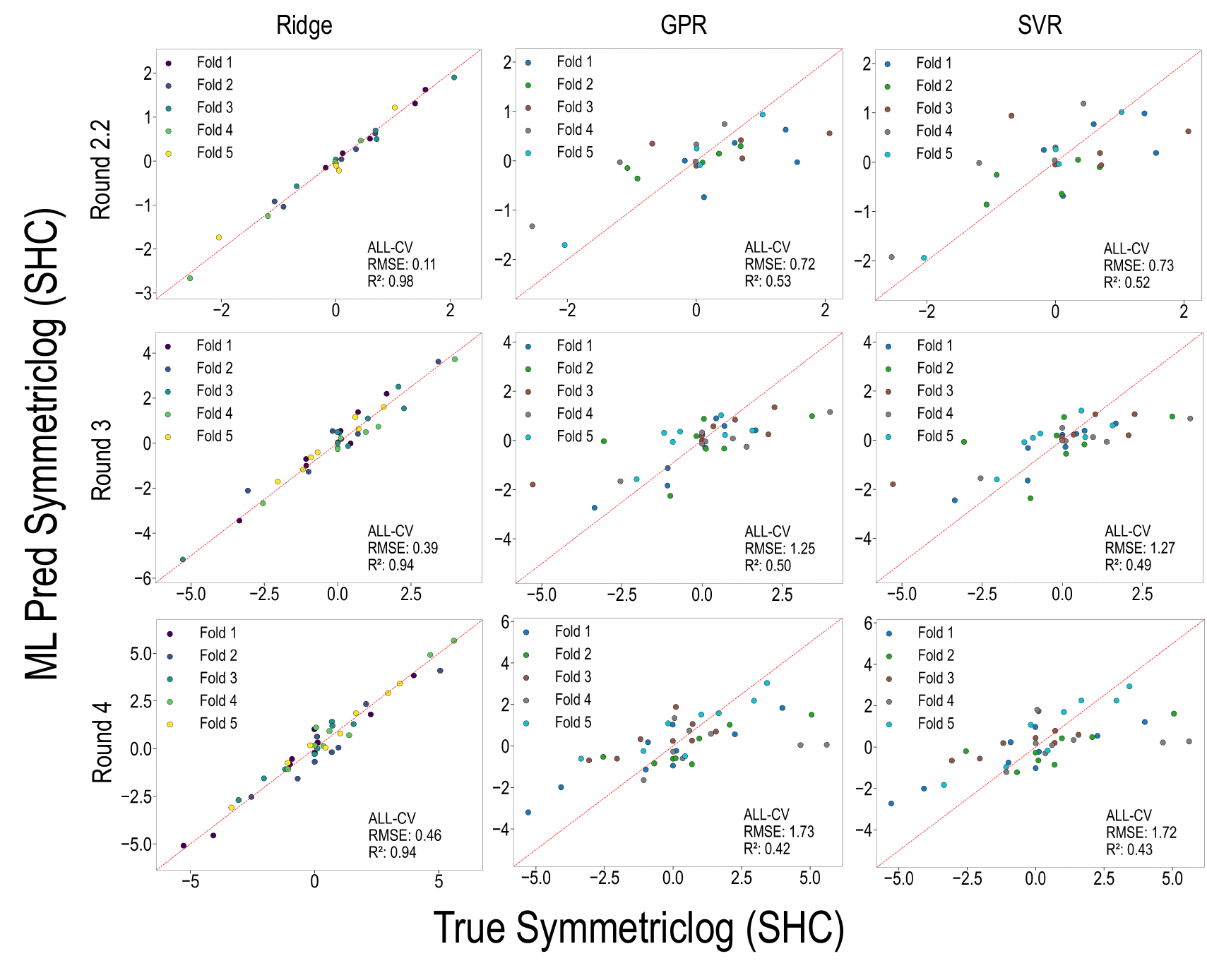}
    \caption{\textbf{Performance of different machine learning (ML) models.} Parity plots indicating performance of different ML models trained on SHC data across different active learning rounds. Results for 3 different regression models, namely, ridge regression (Ridge), Gaussian process regression (GPR) and support vector regression (SVR) are shown, with SHC values plotted in symmetric log space. Colored markers represent particular CV fold, while red dashed line indicate the ideal parity line. Performance metrics for the overall 5-fold CV, including root mean square error (RMSE) and coefficient of determination (R$^2$), are reported. Ridge regression can be seen to consistently outperform other models, achieving lower RMSE and higher R$^2$ values.}
    \label{fig:supple3.1}
\end{figure}

\begin{figure}[ht]
    \centering
    \includegraphics[width=\textwidth]{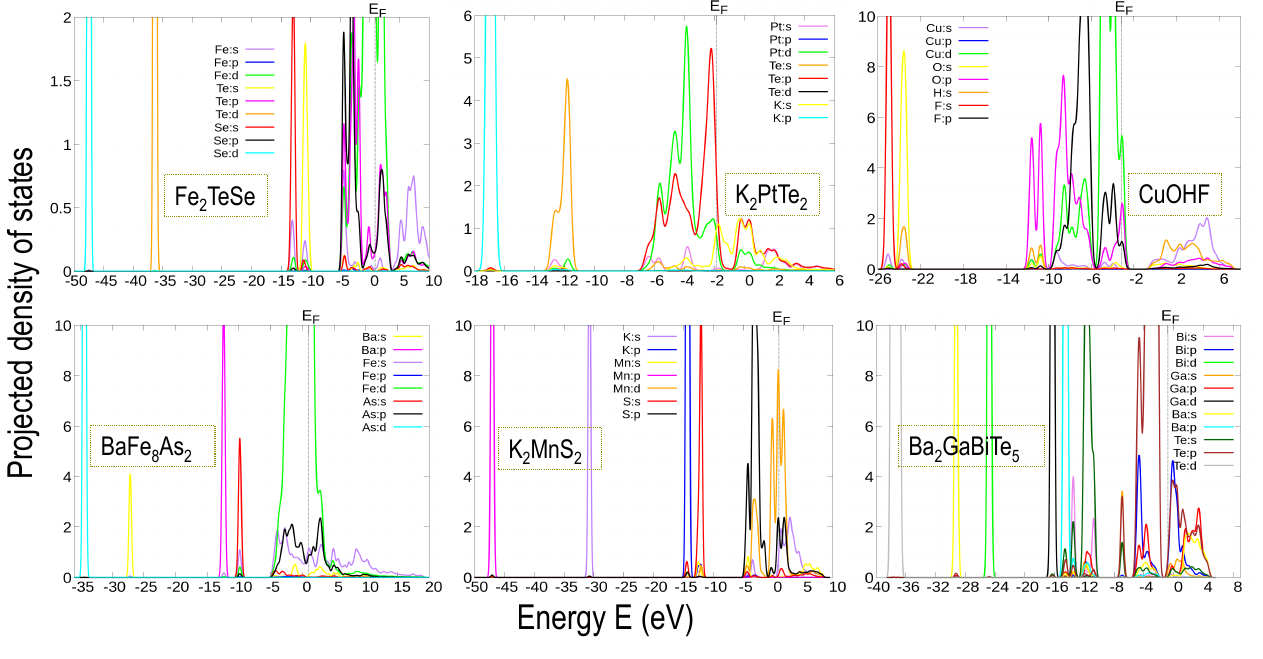}
    \caption{\textbf{Projected density of states (PDOS)} for top 6 candidates showing contribution of their constituent atomic orbitals providing information on appropriate projectors for Wannierization of bands.}
    \label{fig:supple3.1}
\end{figure}

\begin{figure}[ht]
    \centering
    \includegraphics[width=\textwidth]{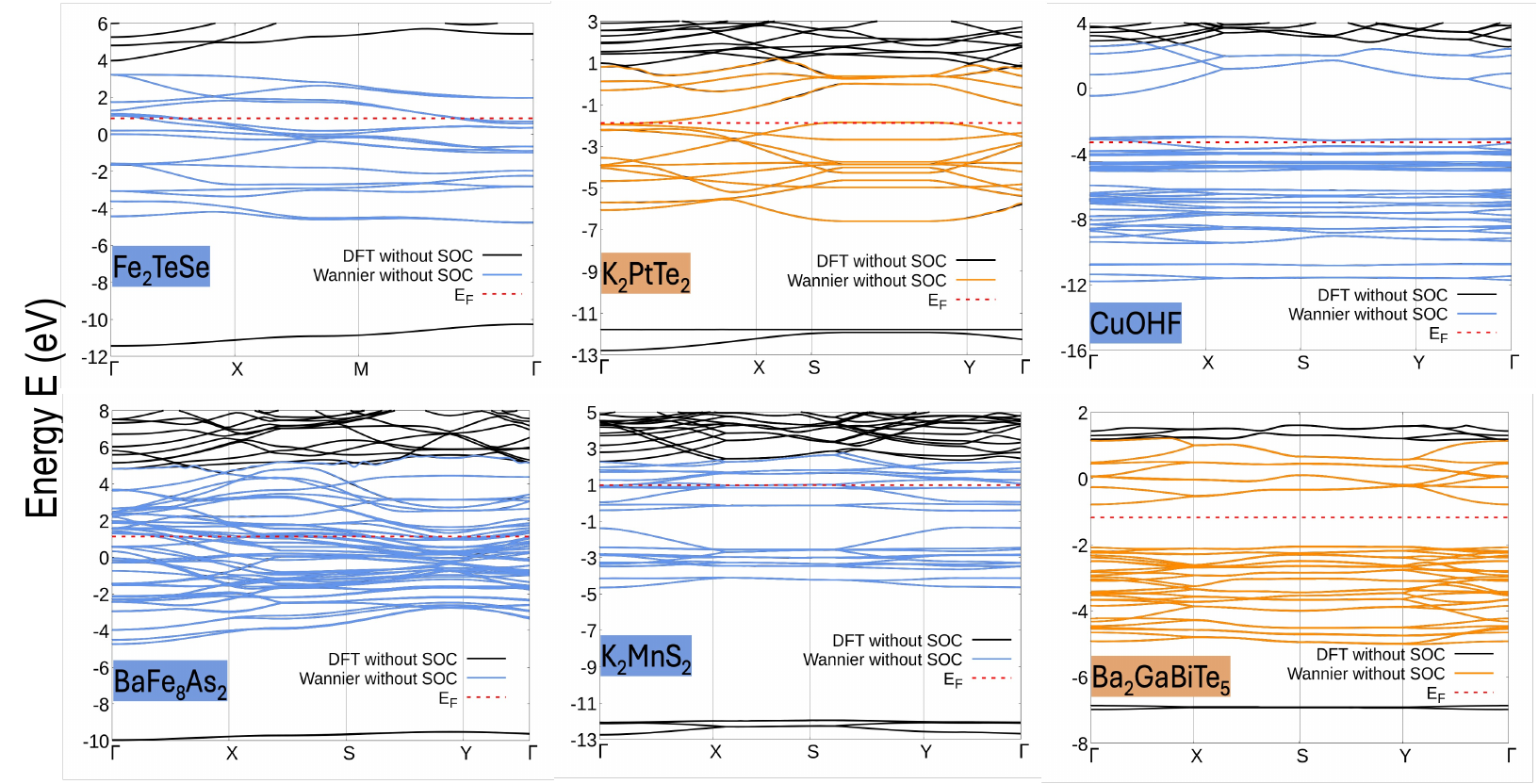}
    \caption{\textbf{MLWF fitted DFT electronic band structures (EBS).} The \textit{ab initio} EBS along high symmetry \textit{k}-path for some top candidates from round 3 (blue) and round 2.2 (orange) fitted with strongly converged MLWFs without spin-orbit coupling (SOC). The SOC corrections are later added to the TB part.}
    \label{fig:supple3.1}
\end{figure}

\begin{figure}[ht]
    \centering
    \includegraphics[width=8.8cm, height=7.5cm]{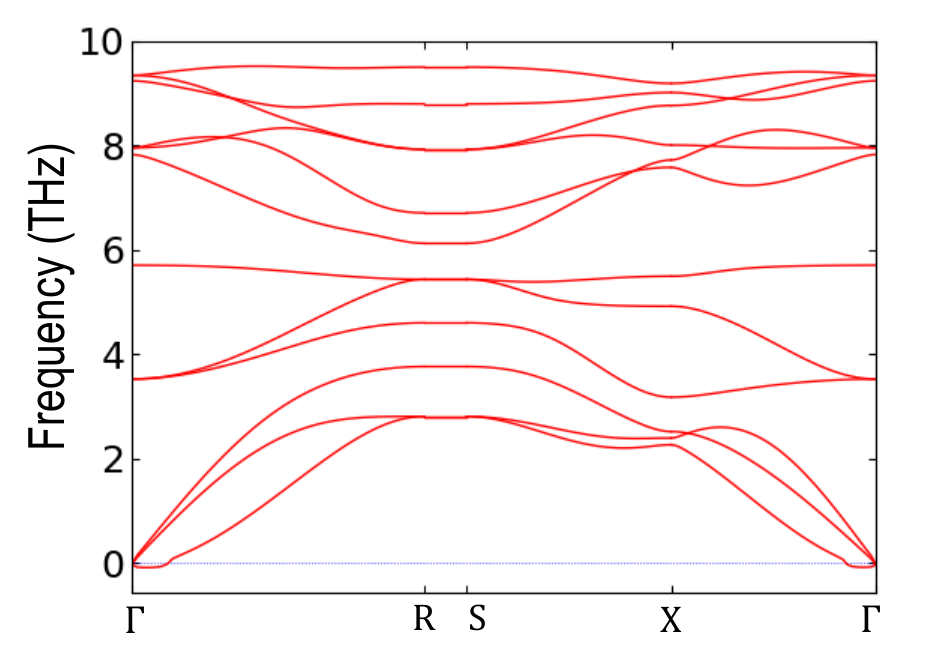}
    \caption{\textcolor{black}{\textbf{Phonon band structure for rank-1 candidate Fe\textsubscript{2}TeSe} showing negligible
    negative frequency of around 0.16 THz indicating the system is dynamically stable.}}
    \label{fig:supple3.1}
\end{figure}

\begin{figure}[!htb]
    \centering
    \includegraphics[scale=0.5]{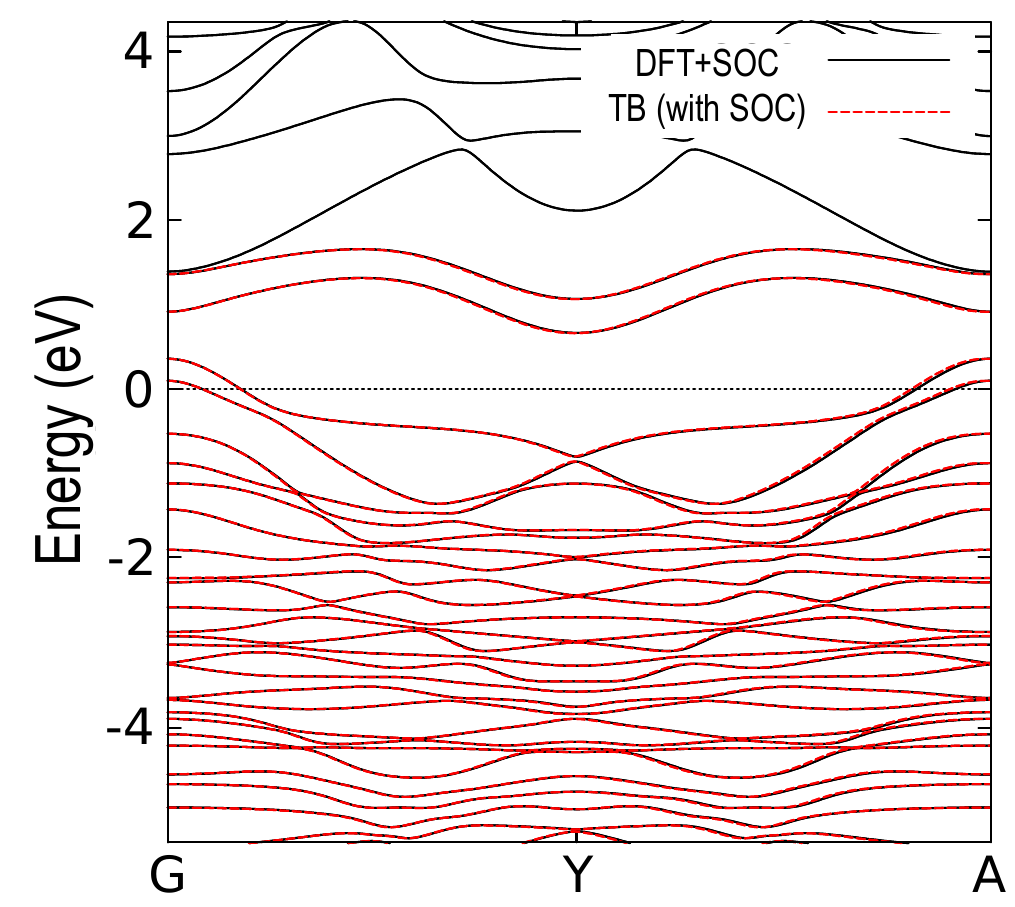}
    \caption{\textcolor{black}{\textbf{Comparison between bands of DFT+SOC and TB (with SOC) for AuSe.} It shows a perfect match indicating correctness of both eigenvalues and eigenstates. The corresponding value of SOC parameters obtained by fitting are $\lambda_{Au,d} = 0.61\ eV$ and $\lambda_{Se,p} = 0.33\ eV$.}}
    \label{fig:placeholder}
\end{figure}

\begin{table}[h]
\vspace{0.5em}
\centering
\noindent\parbox{\textwidth}{%
  \textbf{Table S1.} \fontsize{11}{13}\selectfont \textcolor{black}{{\textbf{Spin-orbit coupling (SOC) parameters $\lambda$}} obtained by fitting the TB bands with DFT(SOC). Note that the parameters are listed  corresponding to the orbital projections used for the given materials. One may notice that some of the lighter elements have a larger SOC than some heavier elements, because of the fact that  p-orbitals usually have larger SOC as compared to the d-orbitals.%
\\}}
%\caption{Spin-orbit coupling parameters($\lambda$) obtained by fitting the TB bands with DFT(SOC). Note that the parameters are listed  corresponding to the orbital projections used for the given materials. One may notice that some of the lighter elements have a larger SOC than some heavier elements, because of the fact that  p-orbitals usually have larger SOC as compared to the d-orbitals.}
\begin{tabular}{l l}
\toprule
\textcolor{black}{Material} & \textcolor{black}{SOC parameters (eV)}\\
\midrule
\textcolor{black}{AuI} & \textcolor{black}{$\lambda_{Au}=0.584$, $\lambda_{I}=0.622$}\\
\textcolor{black}{ZrS$_{2}$} & \textcolor{black}{$\lambda_{Zr}=0.058$, $\lambda_{S}=0.0795$}\\
\textcolor{black}{LiO} & \textcolor{black}{$\lambda_{Li}=0$, $\lambda_{O}=0.0351$}\\
\textcolor{black}{In$_{2}$Cl$_{6}$} & \textcolor{black}{$\lambda_{In}=0$, $\lambda_{Cl}=0.112$}\\
\textcolor{black}{ZrCl$_{2}$} & \textcolor{black}{$\lambda_{Zr}=0.058$, $\lambda_{Cl}=0$}\\
\textcolor{black}{As$_{2}$S$_{3}$} & \textcolor{black}{$\lambda_{As}=0.241$, $\lambda_{S}=0.0712$}\\
\textcolor{black}{ZrNI(Pmmn)} & \textcolor{black}{$\lambda_{Zr}=0.055$, $\lambda_{I}=0.814$, $\lambda_{N}=0.0181$}\\
\textcolor{black}{Na$_{2}$PtO$_{6}$H$_{6}$} & \textcolor{black}{$\lambda_{Pt}=0.568$, $\lambda_{O}=0.0357$, $\lambda_{H}=0$, $\lambda_{Na}=0$}\\
\textcolor{black}{CdI$_{2}$} & \textcolor{black}{$\lambda_{Cd}=0.312$, $\lambda_{I}=0.829$}\\
\textcolor{black}{FeOCl} & \textcolor{black}{$\lambda_{Fe}=0.0691$, $\lambda_{O}=0.0356$, $\lambda_{Cl}=0.121$}\\
\textcolor{black}{PbS$_{2}$} & \textcolor{black}{$\lambda_{Pb}=0$, $\lambda_{S}=0.0795$}\\
\textcolor{black}{BiTe$_{2}$} & \textcolor{black}{$\lambda_{Bi}=1.65$, $\lambda_{Te}=0.74$}\\
\textcolor{black}{ZrTe$_{2}$} & \textcolor{black}{$\lambda_{Zr}=0.059$, $\lambda_{Te}=0.71$}\\
\textcolor{black}{SbAs} & \textcolor{black}{$\lambda_{Sb}=0.545$, $\lambda_{As}=0.246$}\\
\textcolor{black}{Sb$_{2}$Te$_{2}$} & \textcolor{black}{$\lambda_{Sb}=0.529$, $\lambda_{Te}=0.707$}\\
\textcolor{black}{MoS$_{2}$} & \textcolor{black}{$\lambda_{Mo}=0.0712$, $\lambda_{S}=0.0715$}\\
\textcolor{black}{AgN$_{3}$} & \textcolor{black}{$\lambda_{N}=0.0196$, $\lambda_{Ag}=0.271$}\\
\textcolor{black}{SbI$_{3}$} & \textcolor{black}{$\lambda_{Sb}=0.536$, $\lambda_{I}=0.891$}\\
\textcolor{black}{Sb$_{2}$S$_{3}$} & \textcolor{black}{$\lambda_{Sb}=0.541$, $\lambda_{S}=0.0771$}\\
\textcolor{black}{Li$_{2}$CuAs} & \textcolor{black}{$\lambda_{Cu}=0.122$, $\lambda_{S}=0.0797$, $\lambda_{Li}=0$}\\
\textcolor{black}{MoSe$_{2}$} & \textcolor{black}{$\lambda_{Mo}=0.0718$, $\lambda_{Se}=0.346$}\\
\textcolor{black}{PtS$_{2}$} & \textcolor{black}{$\lambda_{Pt}=0.581$, $\lambda_{S}=0.0788$}\\
\textcolor{black}{Li$_{2}$PtH$_{2}$} & \textcolor{black}{$\lambda_{Pt}=0.547$, $\lambda_{Li}=0$, $\lambda_{H}=0$}\\
\textcolor{black}{ZrNI} & \textcolor{black}{$\lambda_{Zr}=0.052$, $\lambda_{I}=0.887$, $\lambda_{N}=0.0152$}\\
\textcolor{black}{MoTe$_{2}$} & \textcolor{black}{$\lambda_{Mo}=0.0912$, $\lambda_{Te}=0.77$}\\
\textcolor{black}{AgBr} & \textcolor{black}{$\lambda_{Ag}=0.242$, $\lambda_{Br}=0.414$}\\
\textcolor{black}{GeSe} & \textcolor{black}{$\lambda_{Ge}=0$, $\lambda_{Se}=0$}\\
\textcolor{black}{PtSe$_{2}$} & \textcolor{black}{$\lambda_{Pt}=0.568$, $\lambda_{Se}=0.351$}\\
\textcolor{black}{PbTe} & \textcolor{black}{$\lambda_{Pb}=1.25$, $\lambda_{Te}=0.69$}\\
\textcolor{black}{WSe$_{2}$} & \textcolor{black}{$\lambda_{W}=0.228$, $\lambda_{Se}=0.364$}\\
\textcolor{black}{AuSe} & \textcolor{black}{$\lambda_{Au}=0.611$, $\lambda_{Se}=0.328$}\\
\textcolor{black}{MnC$_{6}$N$_{4}$Se$_{2}$} & \textcolor{black}{$\lambda_{Mn}=0.0538$, $\lambda_{C}=0.00844$, $\lambda_{Se}=0.312$, $\lambda_{N}=0.0163$}\\
\textcolor{black}{NiTe} & \textcolor{black}{$\lambda_{Ni}=0.084$, $\lambda_{Te}=0.692$}\\
\textcolor{black}{K$_{8}$ZnTl$_{10}$} & \textcolor{black}{$\lambda_{Zn}=0.133$, $\lambda_{Tl}=0.856$, $\lambda_{K}=0$}\\
\textcolor{black}{TaNi$_{2}$TeSe} & \textcolor{black}{$\lambda_{Ta}=0.239$, $\lambda_{Ni}=0.099$, $\lambda_{Te}=0.679$, $\lambda_{Se}=0.357$}\\
\textcolor{black}{Ba$_{2}$GaBiTe$_{5}$} & \textcolor{black}{$\lambda_{Te}=0.712$, $\lambda_{Bi}=1.62$, $\lambda_{Ba}=0$, $\lambda_{Ga}=0$ }\\
\textcolor{black}{K$_{2}$MnS$_{2}$} & \textcolor{black}{$\lambda_{Mn}=0.0546$, $\lambda_{S}=0.0671$, $\lambda_{K}=0$}\\
\textcolor{black}{BaFe$_{8}$As$_{2}$} & \textcolor{black}{$\lambda_{Fe}=0.0697$, $\lambda_{As}=0.233$, $\lambda_{Ba}=0$}\\
\textcolor{black}{CuOHF} & \textcolor{black}{$\lambda_{Cu}=0.121$, $\lambda_{F}=0.059$, $\lambda_{O}=0.0357$, $\lambda_{H}=0$}\\
\textcolor{black}{K$_{2}$PtTe$_{2}$} & \textcolor{black}{$\lambda_{Pt}=0.568$, $\lambda_{Te}=0.706$, $\lambda_{K}=0$}\\
\textcolor{black}{Fe$_{2}$TeSe} & \textcolor{black}{$\lambda_{Fe}=0.0688$, $\lambda_{Se}=0.354$, $\lambda_{Te}=0.719$}\\
\bottomrule
\end{tabular}
\end{table}

\begin{table}[h]
\vspace{5em}
\centering
\noindent\parbox{\textwidth}{
\textbf{Table S2.}
\textbf{A brief comparison between the SHC for few TMD systems} from the literature and the current work, which shows an excellent agreement. All are in the units of $\hbar$/e $\Omega$\textsuperscript{-1}.
}\\[6pt]
\begin{minipage}{\linewidth}
\centering
    \begin{tabular}{c c c}
    \hline\hline
      Systems   & SHC (*) & SHC (This work) \\
      \hline
      MoS$_2$ & 1.0  & 1.0\\
      MoSe$_2$& 1.8& 1.8\\
      MoTe$_2$ & 3.0& 3.0\\
      WSe$_2$ & 7.1 & 8.5\\
      \hline\hline
    \end{tabular}\\[4pt]
\raggedright\footnotesize\hspace{0.9em}\parbox{1.0\linewidth}{* Bhowal, Sayantika, and Sashi Satpathy. ``Intrinsic orbital and spin Hall effects in monolayer transition metal dichalcogenides.'' \textit{Physical Review B} 102.3 (2020): 035409.}
\end{minipage}
\end{table}

\begin{table}[h]
\centering
\vspace{1em}
\noindent\parbox{\textwidth}{
\textbf{Table S3.} %\fontsize{11}{13}\selectfont 
\textbf{ The 26 hand-crafted features used for ML model development.} The remaining list of Magpie features can be found online at \textit{https://hackingmaterials.lbl.gov/matminer/}.%
}
%\caption{{\textbf{The 26 hand-crafted features used for ML model development}. The remaining list of Magpie features can be found online at \textit{https://hackingmaterials.lbl.gov/matminer/}.%
%}}
\vspace{0.5em}

\begin{tabular}{|p{3.7cm}|p{12cm}|}
\hline
\textbf{Acronym} & \textbf{Description} \\
\hline
SG number &	Space group number \\
%\hline
%SG symbol &	Space group symbol \\
\hline
sp\_db\_eg &	MC2D database bandgap \\
%\hline
%Calc.Eg &	DFT calculated bandgap in present study \\
\hline
sp\_rot &	Rotational symmetry  present?- Yes (1) or No (0) \\
\hline
sp\_n\_rot &	Highest order of rotational symmetry \\
\hline
sp\_inv	& Inversion symmetry present? - Yes (1) or No (0) \\
\hline
sp\_rotoinv &	Rotoinversion  symmetry present? - Yes (1) or No (0) \\
\hline
sp\_n\_rotoinv &	Higher order of axis of rotoinversion symmetry \\
\hline
sp\_mir &	Mirror symmetry present? - Yes (1) or No (0) \\
\hline
sp\_n\_mirror &	Number of mirror planes \\
\hline
sp\_screw &	Screw axis symmetry present? - Yes (1) or No (0) \\
\hline
sp\_n\_rot\_screw &	Highest order of rotation in screw symmetry \\
\hline
sp\_1/2tran\_screw &	Half translation in screw symmetry present? - Yes (1) or No (0) \\
\hline
sp\_1/4tran\_screw &	Quarter translation in screw symmetry present? - Yes (1) or No (0) \\
\hline
sp\_glide &	Glide plane symmetry  present? - Yes (1) or No (0) \\
\hline
sp\_1/2tran\_glide &	Half translation in glide symmetry present? - Yes (1) or No (0) \\
\hline
sp\_1/4tran\_glide &	Quarter translation in glide symmetry present? - Yes (1) or No (0) \\
\hline
sp\_val\_metal\_per\_atom	& Number of valence electrons in PBE norm-conserving pseudopotential (NCPP) for metal divided by total atoms present in formula unit \\
\hline
sp\_val\_s\_metal\_per\_atom	& Number of \textit{s}-orbital valence electrons in PBE NCPP for metal divided by total atoms present in formula unit \\
\hline
sp\_val\_p\_metal\_per\_atom	& Number of \textit{p}-orbital valence electrons in PBE NCPP for metal divided by total atoms present in formula unit \\
\hline
sp\_val\_d\_metal\_per\_atom	& Number of \textit{d}-orbital valence electrons in PBE NCPP for metal divided by total atoms present in formula unit \\
\hline 
sp\_val\_f\_metal\_per\_atom	& Number of \textit{f}-orbital valence electrons in PBE NCPP for metal divided by total atoms present in formula unit \\
\hline
sp\_val\_anion\_per\_atom	& Number of valence electrons in PBE NCPP for anions divided by total atoms present in formula unit \\
\hline 
sp\_val\_s\_anion\_per\_atom	& Number of \textit{s}-orbital valence electrons in PBE NCPP for anions divided by total atoms present in formula unit \\
\hline 
sp\_val\_p\_anion\_per\_atom	& Number of \textit{p}-orbital valence electrons in PBE NCPP for anions divided by total atoms present in formula unit \\
\hline
sp\_val\_d\_anion\_per\_atom	& Number of \textit{d}-orbital valence electrons in PBE NCPP for anions divided by total atoms present in formula unit \\
\hline 
sp\_val\_f\_anion\_per\_atom &	Number of \textit{f}-orbital valence electrons in PBE NCPP for anions divided by total atoms present in formula unit \\
%\hline
%sp\_magnetic	& Magnetic state as stated in MC2D, 0 for non-magnetic, 1 for ferromagnetic, -1 for antiferromagentic \\
\hline
\end{tabular}
\end{table}

\begin{table}[h!]
\centering
\vspace{0.1em}
\noindent\parbox{\textwidth}{%
  \textbf{Table S4}: \fontsize{11}{13}\selectfont \textbf{List of features} retained after recursive feature elimination (RFE) across active ML rounds.%
}
\vspace{0.1em}

\begin{tabular}{l l l l} % three columns, all left aligned
\toprule
\textbf{} & \textbf{Round 2.2} & \textbf{Round 3} & \textbf{Round 4} \\
\midrule
 & \textit{Common in all three rounds} & \\
\midrule
1 & \parbox{5cm}{sp\_1/2tran\_screw} & \parbox{5cm}{sp\_1/2tran\_screw} & \parbox{5cm}{sp\_1/2tran\_screw} \\
\\[-5ex]
& & \\
2 & \parbox{5cm}{sp\_1/2tran\_glide} & \parbox{5cm}{sp\_1/2tran\_glide} & \parbox{5cm}{sp\_1/2tran\_glide} \\
\\[-5ex]
& & \\
3 & \parbox{5cm}{MagpieData maximum NValence} & \parbox{5cm}{MagpieData maximum NValence} & \parbox{5cm}{MagpieData maximum NValence} \\
\\[-5ex]
& & \\
\midrule
 & \textit{Common in round 2.2 and 3} & \\
\midrule
\\[-5ex]
& & \\
4 & \parbox{5cm}{MagpieData avg\_dev AtomicWeight} & \parbox{5cm}{MagpieData avg\_dev AtomicWeight} &  \\
\\[-5ex]
& & \\
5 & \parbox{5cm}{MagpieData mean MeltingT} & \parbox{5cm}{MagpieData mean MeltingT} &  \\
\\[-5ex]
& & \\
6 & \parbox{5cm}{MagpieData minimum NValence} & \parbox{5cm}{MagpieData minimum NValence} &  \\
\\[-5ex]
& & \\
7 & \parbox{5cm}{MagpieData mean NdUnfilled} & \parbox{5cm}{MagpieData mean NdUnfilled} & \\
\\[-5ex]
& & \\
\midrule
 & \textit{Common in round 2.2 and 4} & \\
\midrule
\\[-5ex]
& & \\
8 & \parbox{5cm}{MagpieData minimum MeltingT} & - - - - - - - - - - - - - - & \parbox{5cm}{MagpieData minimum MeltingT} \\
\\[-5ex]
& & \\
9 & \parbox{5cm}{MagpieData mean CovalentRadius} & - - - - - - - - - - - - - - & \parbox{5cm}{MagpieData mean CovalentRadius} \\
\\[-5ex]
& & \\
10 & \parbox{5cm}{MagpieData avg\_dev NpValence} & - - - - - - - - - - - - - - & \parbox{5cm}{MagpieData avg\_dev NpValence} \\
\\[-5ex]
& & \\
11 & \parbox{5cm}{MagpieData maximum GSvolume\_pa} & - - - - - - - - - - - - - - & \parbox{5cm}{MagpieData maximum GSvolume\_pa} \\
\\[-5ex]
& & \\
\midrule
 & \textit{Common in round 3 and 4} & \\
\midrule
\\[-5ex]
& & \\
12 & - - - - - - - - - - - - - - & \parbox{5cm}{sp\_val\_p\_metal\_per\_atom} & \parbox{5cm}{sp\_val\_p\_metal\_per\_atom} \\
\\[-5ex]
& & \\
13 & - - - - - - - - - - - - - - & \parbox{5cm}{sp\_val\_d\_anion\_per\_atom} & \parbox{5cm}{sp\_val\_d\_anion\_per\_atom} \\
\\[-5ex]
& & \\
14 & - - - - - - - - - - - - - - & \parbox{5cm}{MagpieData minimum Number} & \parbox{5cm}{MagpieData minimum Number} \\
\\[-5ex]
& & \\
15 & - - - - - - - - - - - - - - & \parbox{5cm}{MagpieData mean Number} & \parbox{5cm}{MagpieData mean Number} \\
\\[-5ex]
& & \\
16 & - - - - - - - - - - - - - - & \parbox{5cm}{MagpieData range CovalentRadius} & \parbox{5cm}{MagpieData range CovalentRadius} \\
\\[-5ex]
& & \\
17 & - - - - - - - - - - - - - - & \parbox{5cm}{MagpieData minimum Electronegativity} & \parbox{5cm}{MagpieData minimum Electronegativity} \\
\\[-5ex]
& & \\
18 & - - - - - - - - - - - - - - & \parbox{5cm}{MagpieData mode Electronegativity} & \parbox{5cm}{MagpieData mode Electronegativity} \\
\\[-5ex]
& & \\
19 & - - - - - - - - - - - - - - & \parbox{5cm}{MagpieData range NdValence} & \parbox{5cm}{MagpieData range NdValence} \\
\\[-5ex]
& & \\
20 & - - - - - - - - - - - - - - & \parbox{5cm}{MagpieData mode NpUnfilled} & \parbox{5cm}{MagpieData mode NpUnfilled} \\
\\[-5ex]
& & \\
21 & - - - - - - - - - - - - - - & \parbox{5cm}{MagpieData mode NdUnfilled} & \parbox{5cm}{MagpieData mode NdUnfilled} \\
\\[-5ex]
& & \\
22 & - - - - - - - - - - - - - - & \parbox{5cm}{MagpieData mean GSmagmom} & \parbox{5cm}{MagpieData mean GSmagmom} \\
\\[-5ex]
& & \\
23 & - - - - - - - - - - - - - - & \parbox{5cm}{MagpieData minimum SpaceGroupNumber} & \parbox{5cm}{MagpieData minimum SpaceGroupNumber} \\
\\[-5ex]
& & \\
24 & - - - - - - - - - - - - - - & \parbox{5cm}{MagpieData maximum SpaceGroupNumber} & \parbox{5cm}{MagpieData maximum SpaceGroupNumber} \\
\\[-5ex]
& & \\
25 & - - - - - - - - - - - - - - & \parbox{5cm}{sp\_inv} & \parbox{5cm}{sp\_inv} \\
\\[-5ex]
& & \\
\midrule
\end{tabular}
\label{tab:simple-table}
\end{table}

\begin{table}[h!]
\centering
\begin{tabular}{l l l l} % three columns, all left aligned
\toprule
\textbf{} & \textbf{Round 2.2} & \textbf{Round 3} & \textbf{Round 4} \\
\midrule
 & \textit{Not Common in any rounds} & \\
\midrule
26 & \parbox{5cm}{sp\_rotoinv} & \parbox{5cm}{sp\_val\_metal\_per\_atom} & \parbox{5cm}{sp\_n\_rot} \\
\\[-5ex]
& & \\
27 & \parbox{5cm}{sp\_n\_mirror} & \parbox{5cm}{sp\_val\_d\_metal\_per\_atom} & \parbox{5cm}{MagpieData range Number} \\
\\[-5ex]
& & \\
28 & \parbox{5cm}{MagpieData mode Number} & \parbox{5cm}{MagpieData mode MeltingT} & \parbox{5cm}{MagpieData avg\_dev Number} \\
\\[-5ex]
& & \\
29 & \parbox{5cm}{MagpieData mode AtomicWeight} & \parbox{5cm}{MagpieData range Column} & \parbox{5cm}{MagpieData avg\_dev Number} \\
\\[-5ex]
& & \\
30 & \parbox{5cm}{MagpieData maximum CovalentRadius} & \parbox{5cm}{MagpieData minimum NsValence} & \parbox{5cm}{MagpieData maximum Column} \\
\\[-5ex]
& & \\
31 & \parbox{5cm}{MagpieData mode CovalentRadius} & \parbox{5cm}{MagpieData maximum NsValence} & \parbox{5cm}{MagpieData avg\_dev Row} \\
\\[-5ex]
& & \\
32 & \parbox{5cm}{MagpieData mean NsValence} & \parbox{5cm}{MagpieData avg\_dev NdValence} & \parbox{5cm}{MagpieData mode NpValence} \\
\\[-5ex]
& & \\
33 & \parbox{5cm}{MagpieData mean NdValence} & \parbox{5cm}{MagpieData maximum NsUnfilled} & \parbox{5cm}{MagpieData minimum NdValence} \\
\\[-5ex]
& & \\
34 & \parbox{5cm}{MagpieData mode NdValence} & \parbox{5cm}{MagpieData minimum NdUnfilled} & \parbox{5cm}{MagpieData range NsUnfilled} \\
\\[-5ex]
& & \\
35 & \parbox{5cm}{MagpieData range NValence} & \parbox{5cm}{MagpieData minimum NfUnfilled} & \parbox{5cm}{MagpieData mode NsUnfilled} \\
\\[-5ex]
& & \\
36 & \parbox{5cm}{MagpieData mode NValence} & \parbox{5cm}{MagpieData maximum NfUnfilled} & \parbox{5cm}{MagpieData avg\_dev NpUnfilled} \\
\\[-5ex]
& & \\
37 & \parbox{5cm}{MagpieData mean NpUnfilled} & \parbox{5cm}{MagpieData range NfUnfilled} & \parbox{5cm}{MagpieData range NdUnfilled} \\
\\[-5ex]
& & \\
38 & \parbox{5cm}{MagpieData avg\_dev SpaceGroupNumber} & \parbox{5cm}{MagpieData mean NfUnfilled} & \parbox{5cm}{MagpieData avg\_dev NdUnfilled} \\
\\[-5ex]
& & \\
39 & - - - - - - - - - - - - - - & \parbox{5cm}{MagpieData avg\_dev NfUnfilled} & \parbox{5cm}{MagpieData mode NUnfilled} \\
\\[-5ex]
& & \\
40 & - - - - - - - - - - - - - - & \parbox{5cm}{MagpieData mode NfUnfilled} & \parbox{5cm}{MagpieData avg\_dev GSvolume\_pa} \\
\\[-5ex]
& & \\
41 & - - - - - - - - - - - - - - & \parbox{5cm}{MagpieData range GSvolume\_pa} & \parbox{5cm}{MagpieData maximum GSbandgap} \\
\\[-5ex]
& & \\
42 & - - - - - - - - - - - - - - & \parbox{5cm}{MagpieData range GSbandgap} & \parbox{5cm}{MagpieData mode GSbandgap} \\
\\[-5ex]
& & \\
43 & - - - - - - - - - - - - - - & \parbox{5cm}{MagpieData minimum GSmagmom} & \parbox{5cm}{MagpieData mode GSmagmom} \\
\\[-5ex]
& & \\
44 & - - - - - - - - - - - - - - & \parbox{5cm}{MagpieData maximum GSmagmom} & \parbox{5cm}{MagpieData mean SpaceGroupNumber} \\
\\[-5ex]
& & \\
45 & - - - - - - - - - - - - - - & \parbox{5cm}{MagpieData range GSmagmom} & - - - - - - - - - - - - - -  \\
\\[-5ex]
& & \\
46 & - - - - - - - - - - - - - - & \parbox{5cm}{sp\_1/4tran\_screw} & - - - - - - - - - - - - - -  \\
\\[-5ex]
& & \\
47 & - - - - - - - - - - - - - - & \parbox{5cm}{sp\_1/4tran\_glide} & - - - - - - - - - - - - - -  \\
\\[-5ex]
& & \\
\bottomrule
\end{tabular}
\label{tab:simple-table}
\end{table}